\documentclass[twocolumn,prX,nofootinbib,amsfonts,showpacs,epsfig]{revtex4}%,floats,floatfix,letterpaper]{revtex4}

\usepackage{graphicx}% Include figure files
\usepackage{dcolumn}% Align table columns on decimal point
\usepackage{bm}% bold math
\usepackage{longtable}
\usepackage{amsmath}
\usepackage{mathrsfs}
\usepackage{amsfonts}
\usepackage[usenames]{color}

\usepackage{epstopdf}
\newcommand{\be} {\begin{equation}}
\newcommand{\ee} {\end{equation}}

\newcommand{\bea}{\begin{eqnarray}}
\newcommand{\eea}{\end{eqnarray}}
\newcommand{\bdm}{\begin{displaymath}}
\newcommand{\edm}{\end{displaymath}}
\newcommand{\ba} {\begin{array}}
\newcommand{\ea} {\end{array}}

\newcommand{\bfg}  {\begin{figure}}
\newcommand{\efg}  {\end{figure}}
\newcommand{\bfgd}  {\begin{figure*}}
\newcommand{\efgd}  {\end{figure*}}
\newcommand{\incgr} {\includegraphics}
\newcommand{\la} {\langle}
\newcommand{\ra} {\rangle}
\newcommand{\mbf}{\mathbf}

\newcommand{\btb}{\begin{table}}
\newcommand{\etb}{\end{table}}

\begin{document}

\bibliographystyle{apsrev}

\title{Needlet Detection of Features in WMAP CMB Sky and the Impact on Anisotropies and Hemispherical Asymmetries}
\author{Davide Pietrobon$^{1,2,3}$, Alexandre Amblard$^3$, Amedeo Balbi$^{1,4}$,  Paolo Cabella$^1$,
Asantha Cooray$^3$, Domenico Marinucci$^5$}
\affiliation{
$^1$Dipartimento di Fisica, Universit\`a di Roma ``Tor Vergata'',
via della Ricerca Scientifica 1, 00133 Roma, Italy\\
$^2$Institute of Cosmology and Gravitation, University
of Portsmouth, Mercantile House, Portsmouth PO1 2EG, United Kingdom\\
$^3$Center for Cosmology, University of California, Irvine, California 92697, USA\\
$^4$INFN Sezione di Roma ``Tor Vergata'',
via della Ricerca Scientifica 1, 00133 Roma, Italy\\
$^5$Dipartimento di Matematica, Universit\`a di Roma ``Tor Vergata'',
via della Ricerca Scientifica 1, 00133 Roma, Italy
}

\date{\today}

\begin{abstract}

We apply spherical needlets to the Wilkinson Microwave Anisotropy
Probe 5-year cosmic microwave background (CMB) data\_set, to  search
for imprints of nonisotropic features in the CMB sky. We use the
needlets' localization properties to resolve peculiar features in the
CMB sky and to study how these features contribute to the anisotropy
power spectrum of the CMB. In addition to the now well-known ``cold
spot'' of the CMB map in the southern hemisphere, we also find two
hot spots at greater than 99\% confidence level, again in the
southern hemisphere and closer to the Galactic plane. While the cold
spot contributes to the anisotropy power spectrum  in the multipoles
between $\ell=6$ to $\ell=33$, the hot spots  are found to be
dominating the anisotropy power in the range between $\ell=6$ and
$\ell=18$. Masking both the cold and the two hot spots results in a
reduction by about 15\% in the amplitude of the angular power spectrum of CMB
around $\ell=10$. The resulting changes to the
cosmological parameters when the power spectrum is estimated
masking these features (in addition to the WMAP team's KQ85 mask) are
within the 1$\sigma$ errors published with the WMAP mask only. We also study the
asymmetry between the angular power spectra evaluated on the northern and southern hemispheres.
When the features detected by needlets are masked,
we find that the difference in the power, measured in terms of the
anisotropy variance between $\ell=4$ and $\ell=18$, is reduced by a
factor $2$. We make available a mask related to needlet features for more detailed
studies on asymmetries in the CMB anisotropy sky.

\end{abstract}


\maketitle

% *******************************************************************************************

\section{Introduction}

Beyond the angular power spectrum of the cosmic microwave background (CMB) anisotropies,
the high-sensitivity all-sky CMB maps produced by the Wilkinson Microwave
Anisotropy Probe (WMAP)\footnote{http://lambda.gsfc.nasa.gov/product/map/current/} \cite{WMAP52008}
have enabled detailed statistics studies to  extract higher-order information.
These studies include characterizations of
asymmetries in the CMB sky \cite{Eriksen2008} and the search for anomalous features in the anisotropy field.
Previous analyses have shown evidences for alignments \cite{Copi2007},
asymmetry in the CMB statistics between northern and southern Galactic hemispheres \cite{Eriksen2007},
and features such as the ``cold spot,'' a significant negative feature in the CMB map
first identified with wavelets \cite{Cruz2005ColdSpot}.

Wavelets can be constructed both in real space on the sphere, for
instance by means of the so-called stereographic projection
\cite{AntoineVandergheynst1999,Wiaux2005,McEwen2006b,McEwen2007},
and in harmonic space, for instance following the prescription
described in \cite{Holschneider1996,Mcewen2006a}, see also
\cite{Wiaux-st}. There have been already several applications of
wavelets to CMB data analysis, including tests for non-Gaussianity
and asymmetries
\cite{Vielva2004,Cabella2004,Mcewen2008,Wiaux-ng,Wiaux-glo},
polarization analysis \cite{CabellaNatoliSilk2007}, foreground
subtraction \cite{Hansen2006}, foreground component separation
\cite{Moudden2005,Starck2005}, and point source detection in CMB
anisotropy maps \cite{Sanz2006}. The reason for such a wide success
can be motivated as follows: the comparison between models and CMB
data are primarily undertaken in the Fourier domain, where each
multipole can be measured separately with all-sky experiments.
However, in the presence of sky cuts data analysis must take into
account the coupling among multipoles induced by the mask;
statistical studies consequently become much more challenging.
Wavelets address these issues by providing the possibility to
combine sharp localization properties in pixel space with a peaked
window function in the harmonic space.

In the latest two years, needlets have been proposed as a new
wavelet system which enjoys several advantages over existing
constructions. The introduction of needlets into the mathematical
literature is due to \cite{NarcowichPetrushevWard2006}, where their
localization properties in pixel and harmonic space are established.
The analysis of their statistical properties for spherical random
fields is first provided by \cite{Baldi2006}, where uncorrelation is
discussed and it is also shown how needlets can be used to implement
estimators for the angular power spectrum or tests of
non-Gaussianity. The first application to WMAP data is due to
\cite{Pietrobon2006}, where needlets are used to estimate
(cross-)angular power spectra in order to search for dark energy
imprints on the correlation between large-scale structures and CMB.
A full description of needlets and their potentiality for CMB data
analysis is then provided by \cite{Marinucci2008};
Ref.~\cite{Guilloux2007} investigates the effect of different window
functions in their constructions, whereas Ref.~\cite{Baldi2007}
provides further mathematical results on their behavior for
partially observed sky-maps. The previous results have been extended
to angular power spectrum estimation in the presence of noise
\cite{Fay2008,Fay2008b}, estimation of the bispectrum
\cite{Lan2008NeedBis}, foreground component separation
\cite{Delabrouille2008maps}, and analysis of directional data
\cite{Baldi2008Adaptivedensity}; see also
Refs.~\cite{Geller2007,Lan2008,Mayeli2008} for further developments.

Among the advantages of needlets compared to previously adopted
wavelets, we recall that they are compactly supported in harmonic
space, so that they allow to focus the analysis on a specific set of
multipoles, which can be completely controlled
\cite{NarcowichPetrushevWard2006}. Needlet coefficients can be shown
to be (approximately) uncorrelated both across different frequencies
and, at high multipoles, at different locations in pixel space,
which makes statistical analysis much more efficient; it is possible
to provide an exact analytic expression for their variance and
correlation, in terms of the underlying angular power spectrum
\cite{Baldi2006,Pietrobon2006}. Needlets do not rely on any tangent
plane approximation, but they are directly embedded into the
manifold structure of the sphere; they allow for direct
reconstruction formulas (a consequence of the fact that they make up
a tight frame system) and they are computationally very convenient
\cite{Marinucci2008}.

In this paper, we make use of needlets to further study features in
the WMAP CMB maps. We focus on the large angular scales or,
equivalently, on multipoles smaller than 200. We recover the cold
spot that was previously detected in WMAP data with wavelets, and we
also detect other features, including two hot spots, which have so far
received less attention. By masking these features, we study how the
angular power spectrum of CMB anisotropies is modified. Given that
these features are located in the southern hemisphere, we also
discuss the extent to which these features could be responsible for
the north-south asymmetry in WMAP data
\cite{Lew2008,Eriksen2008,Hansen2004Asym}. As is well known, this
asymmetry has also drawn much interest in the theoretical community,
since it could entail strong implications on the physical nature of
primordial perturbations, including inflation \cite{Erickcek2008}.
By masking the low-$\ell$ features, we find that the difference in
the CMB anisotropy variance between the two hemispheres is reduced
by a factor of 2, reducing the significance of previous
detections.

We also explore the evidence that
statistically significant bumps and dips in the CMB anisotropy power
spectrum at multipoles of 20 and 40 could be related to features in
the CMB sky. We did not locate any particular feature in the sky
that dominates the power spectrum at these multipoles; masking the
significant features detected by needlets tuned to these multipoles
did not change the power spectrum more than $\sim$5\%.

The paper is organized as follows: In the following section we describe
briefly the needlet formalism, then we apply it to the WMAP 5-year
temperature map describing the features we
observed and their significance in Sec.~\ref{sec:gauss_check}. In Sec.~\ref{sec:cls}
we address the angular power spectrum modification induced
by the specific features we measure in the temperature map,
computing the effect on the cosmological parameters.
We then draw our conclusion in Sec.~\ref{sec:conclusions}.

% *******************************************************************************************
\section{Needlets and Search for Features}

We provide here a brief summary of the needlet construction, while
we refer to \cite{Marinucci2008} and the references therein for a
more complete discussion. We recall the spherical needlet function
is defined as

%---------------------------------
\be
  \label{eq:need_def}
  \psi _{jk}(\hat\gamma) = \sqrt{\lambda_{jk}}
  \sum_{\ell} b\Big(\frac{\ell}{B^{j}}\Big)
  \sum_{m=-\ell}^{\ell}\overline{Y}_{\ell m}(\hat\gamma)Y_{\ell m}(\xi _{jk})
\ee
%---------------------------------
where $\{\xi_{jk}\}$ are the cubature points on the sphere,
corresponding to the frequency $j$ and the location $k$, and $b(.)$ is a filter function
in harmonic space. Each cubature point requires a proper weight,
$\lambda_{jk}$. Since our implementation is based on the Healpix
pixelization\footnote{http://healpix.jpl.nasa.gov}
\cite{Gorski2005}, we actually compute Eq.~\ref{eq:need_def} on the
center of the pixels for a given resolution, following the procedure
described in \cite{Pietrobon2006}. In this context the cubature
weights can be approximated by $1/N_p$, where $N_p$ is the number of
pixels for the chosen Healpix resolution and k corresponds to a Healpix pixel number.

In general, needlets can be viewed as a convolution of the
projection operator $\sum_m\overline{Y}_{\ell m}(\hat\gamma)Y_{\ell
m}(\xi _{jk})$ with suitably chosen weights provided by the function
$b(.)$. Details on the function $b(.)$ we used can be found in
Ref.~\cite{Marinucci2008}. We show its profile in Fig.~\ref{fig:b_func}
 for several values of $B^j$; in the sequel we will
use the notation $b_\ell$ for $b\Big(\frac{\ell}{B^{j}}\Big)$. $B$
is a user-chosen parameter that characterizes the weights $b_\ell$
and then the entire set of needlets, since its value determines the
width of the filter function. The choice of $B$ must be driven by
the insight on the range of multipoles to be probed. Since we are
interested in large angular scales, we set $B=1.8$: this choice
allows us to have eight frequencies spanning multipoles up to
$\ell=200$, while the information at smallest scales is concentrated
on just three frequencies. The range of multipoles covered by each
needlet is summarized in Table \ref{tab:l-range}.

\bfg
\incgr[width=\columnwidth]{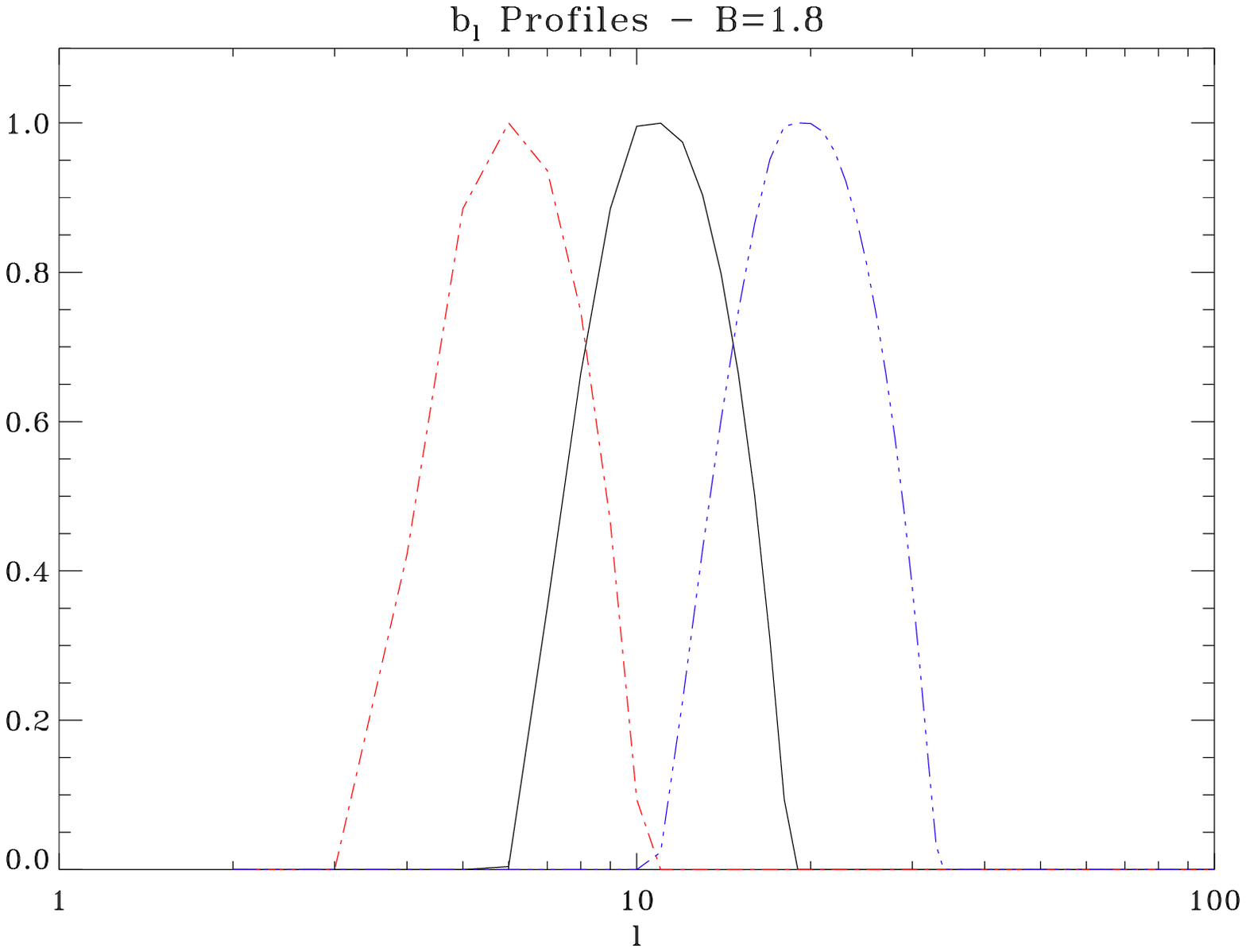}
\incgr[width=\columnwidth]{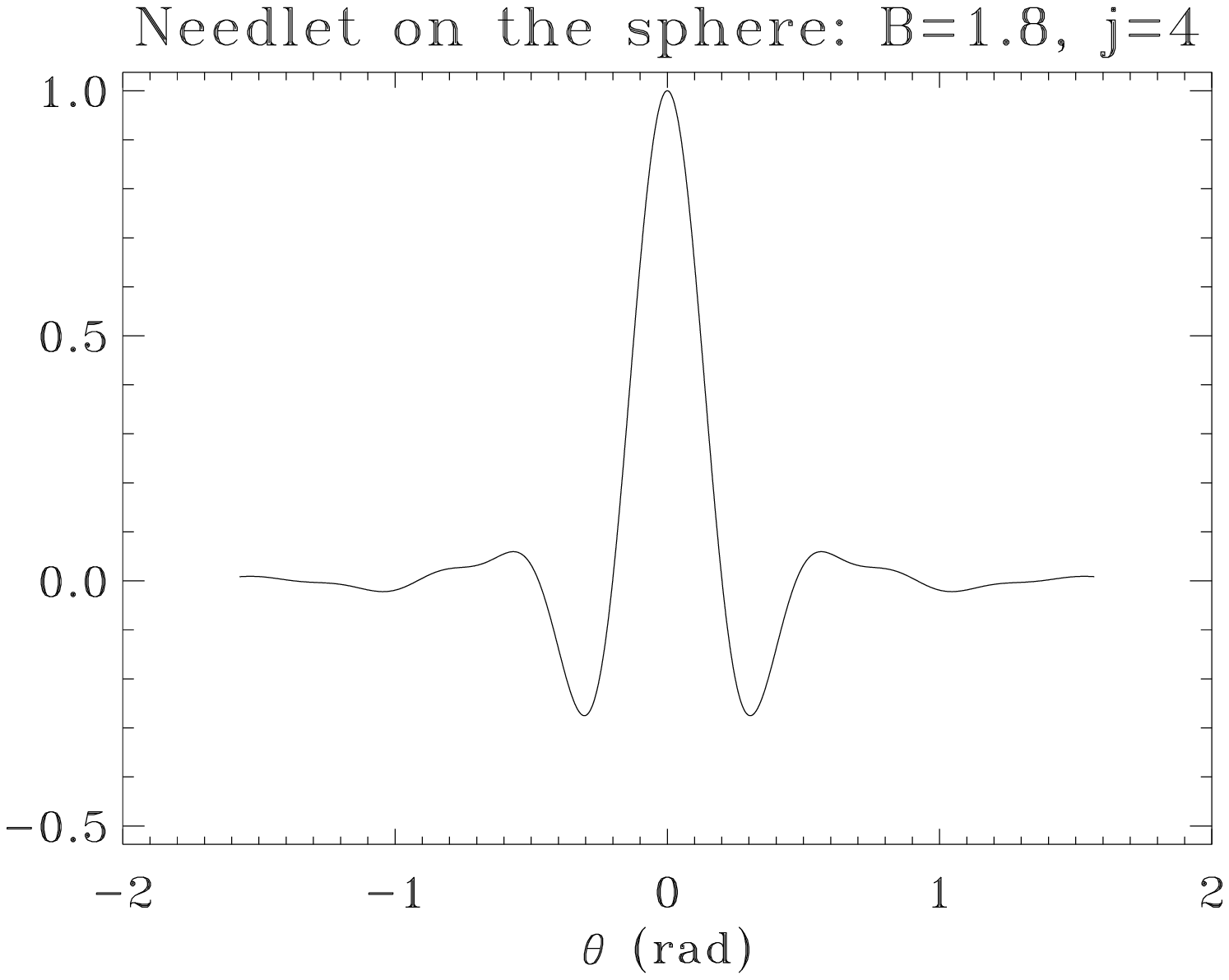}
\caption{Top: Profile of the function $b(x)$ in $\ell$-space for the choice $B=1.8$. The black solid line represents $j=4$, the red dot-dashed line $j=3$ and the blue dot-long-dashed line $j=5$. Bottom: needlets on the sphere for $j=4$. At each cubature point, $\xi_{jk}$, the needlet is sharply localized.}
\label{fig:b_func}
\efg

Since the function $b_\ell$ has a finite support in $\ell$-space, by
construction each needlet receives a contribution only from a
specific range of multipoles, which is related to the index $j$ by a
logarithmic function of $B$. In pixel space needlets are
quasiexponentially localized, due to peculiar properties of the
window function $b_\ell$, as shown by
\cite{NarcowichPetrushevWard2006}. We are thus able to exploit a
tight control on the localization both in terms of angular scales
and directions on the sky, thus keeping a record of the signal in
both domains. This makes possible the detection of specific features
in the CMB map that are contributing power, for example, over a
well-defined range of multipoles of the angular power spectrum.

To this aim, we first decompose the CMB temperature map onto the
needlet system. As recalled before, needlets do not make up a
basis, but a tight frame; however the latter enjoys all the
reconstruction properties that are usually associated to bases
(indeed tight frames can be simply viewed as a redundant basis). The
needlet coefficients $\beta_{jk}$ result from the projection of the
temperature field on the elements of the needlet system:
% --------------------------------
\bea
  \label{eq:need_coef}
  \beta _{jk} &=&\int_{S^{2}}T(\hat\gamma)\psi _{jk}(\hat\gamma)d\Omega  \nonumber \\
   &=&\sqrt{\lambda_{jk}}\sum_{\ell}b\Big(\frac{\ell}{B^{j}}\Big)\sum_{m=-\ell}^{\ell}a_{\ell m}Y_{\ell m}(\xi _{jk}).
\eea
% --------------------------------

\btb[htb]
\begin{center}
\begin{tabular}{|c|c|}
\hline
  j & $\ell$-range \\
\hline
  1 & 2        \\
  2 & 2-5      \\
  3 & 4-10     \\
  4 & 6-18     \\
  5 & 11-33    \\
  6 & 20-60    \\
  7 & 35-108   \\
  8 & 63-196   \\
  9 & 113-352  \\
 10 & 203-635  \\
 11 & 365-1143 \\
\hline
\end{tabular}
\caption{Range of multipoles spanned by needlets for $B=1.8$.}
\label{tab:l-range}
\end{center}
\etb

% *******************************************************************************************
\section{Maps}
\label{sec:w5_need}

We decomposed the Internal Linear Combination (ILC) WMAP 5-year
temperature map by means of needlet functions corresponding to $B=1.8$. As
discussed in the previous section, this specific choice provides us
needlets at 11 frequencies $j$, that span properly the low
multipoles we are interested in. As described above, the expression
\ref{eq:need_coef} can be easily represented by a map in the Healpix
scheme, in which each pixel is associated with the needlet coefficient
evaluated at $\xi_{jk}$. Figure \ref{fig:need_coeff} shows three
clarifying examples of the WMAP 5-year ILC map decomposition on
$j=3$, $j=4$ and $j=5$ needlets by making use of the functions
$b_\ell$ depicted in Fig.~\ref{fig:b_func}.

We first applied our procedure to the WMAP map with the extended
mask KQ$75$ from the WMAP team. Such a large mask, however, entails
a greater correlation for multipoles corresponding to large angular
scales; for low values of $j$, this could potentially impact the
needlets' coefficients as well. Any detection of significant features
in the CMB anisotropy map using the large mask, however, can be
reconfirmed at a higher significance level with a smaller mask.
After studying the temperature map using KQ$75$, we repeated all the
analysis applying KQ$85$, which is the mask currently favored for
cosmological data analysis, including power spectrum measurements.
We find that our results related to statistically significant
features on the WMAP map are fully consistent in the two cases.

\subsection*{III.A  Hot/Cold spot maps}

\begin{figure*}[htb]
\incgr[width=0.6\columnwidth, angle=90]{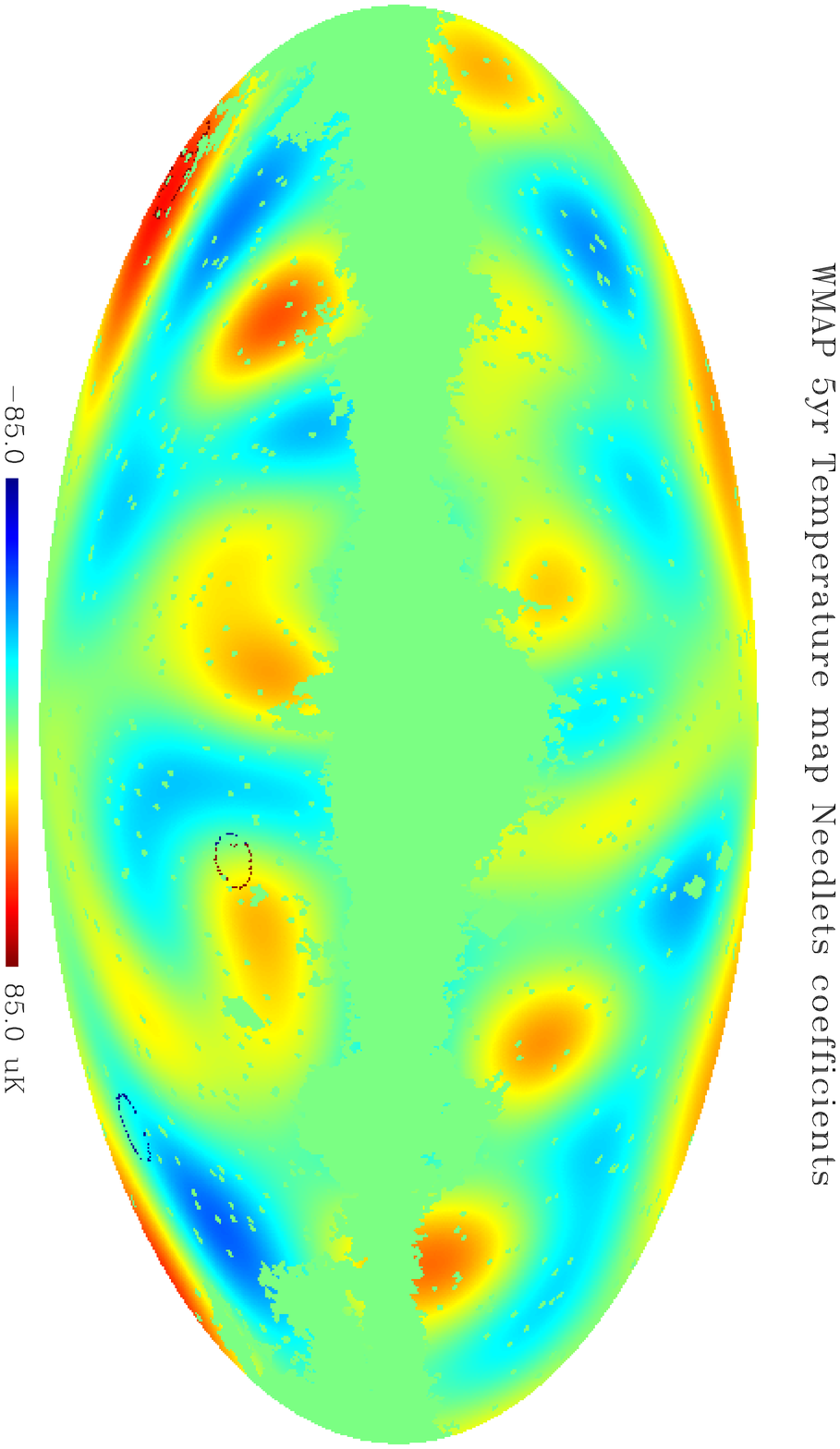}
\incgr[width=0.6\columnwidth, angle=90]{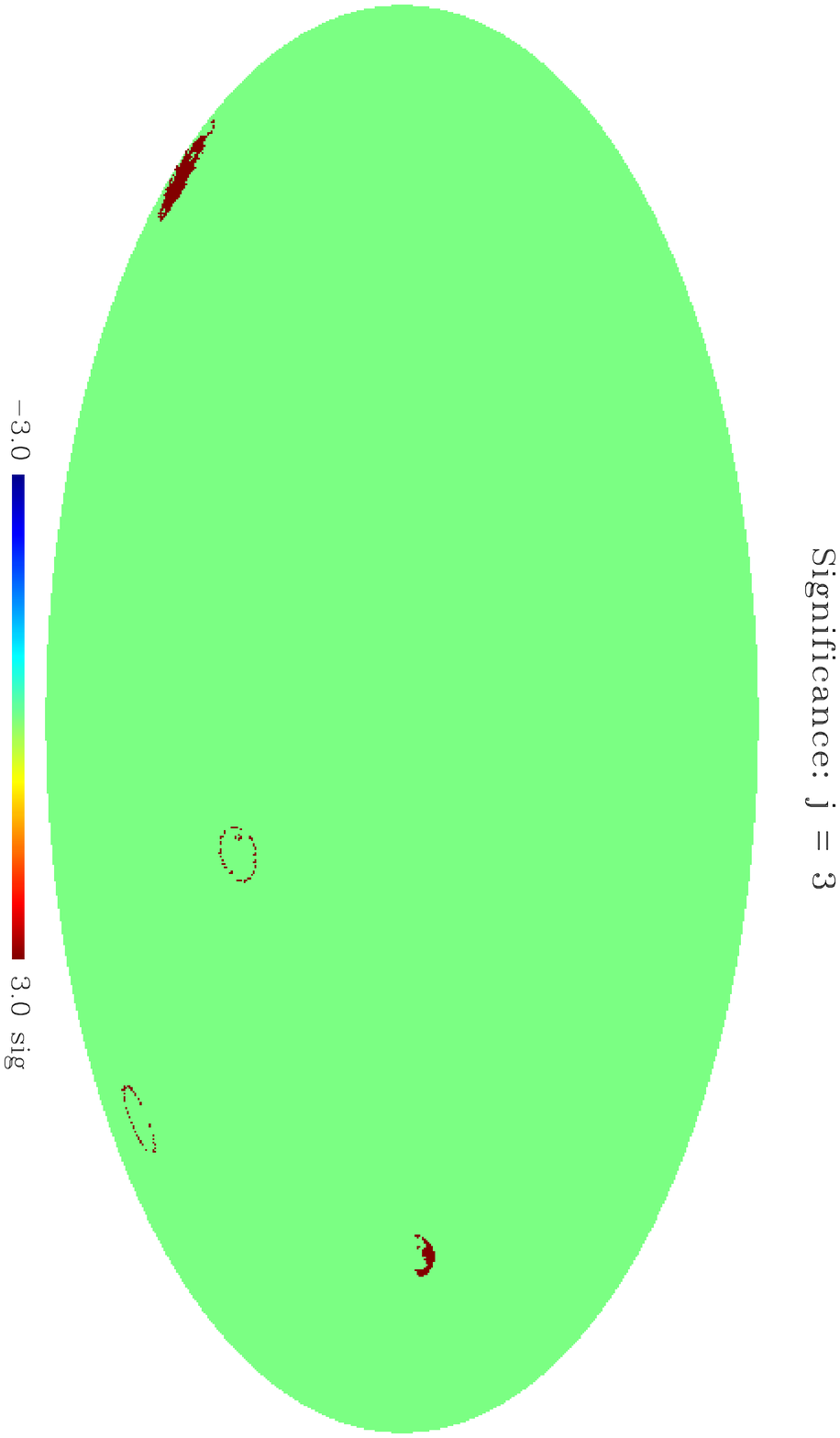}
\incgr[width=0.6\columnwidth, angle=90]{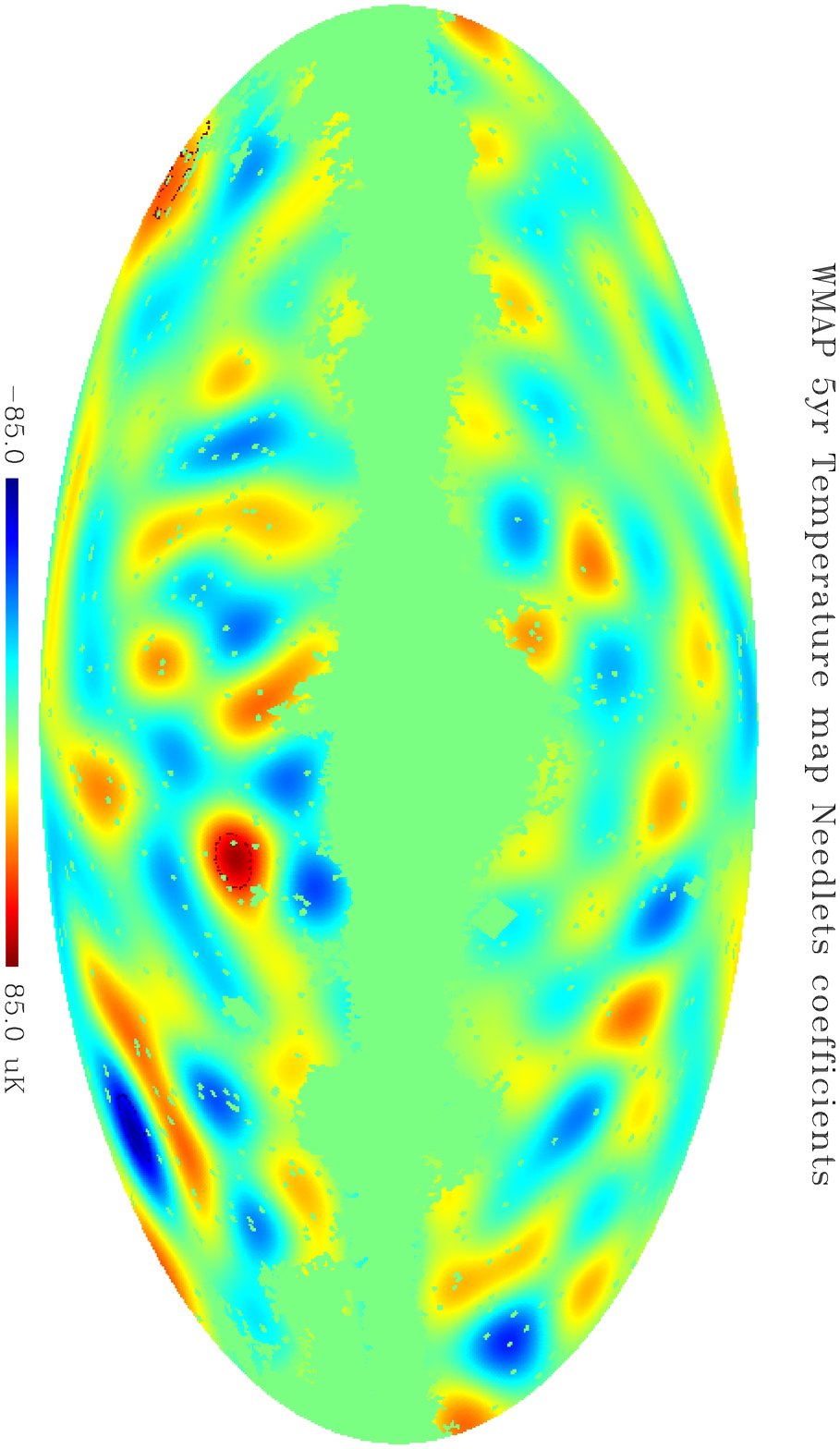}
\incgr[width=0.6\columnwidth, angle=90]{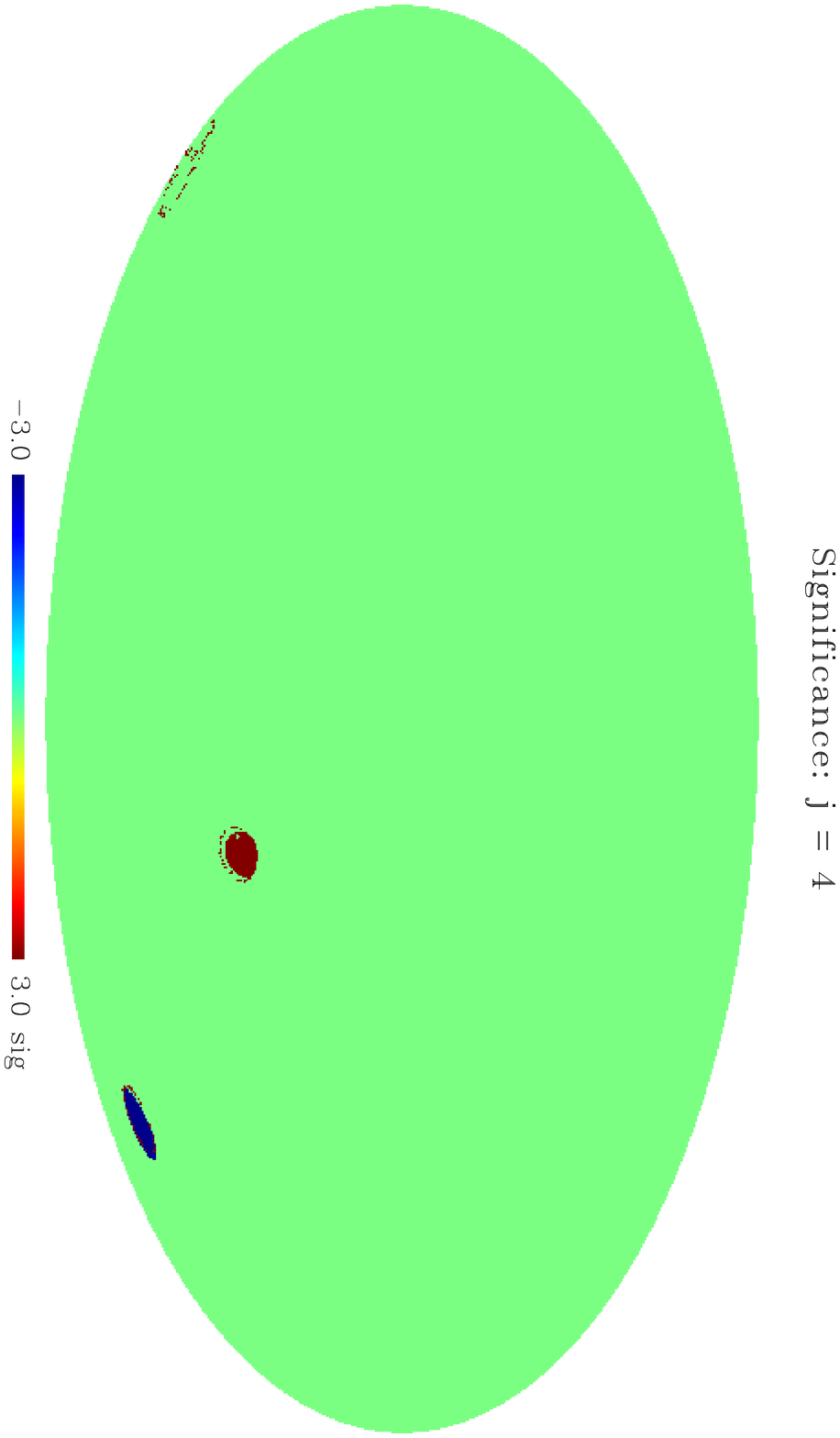}
\incgr[width=0.6\columnwidth, angle=90]{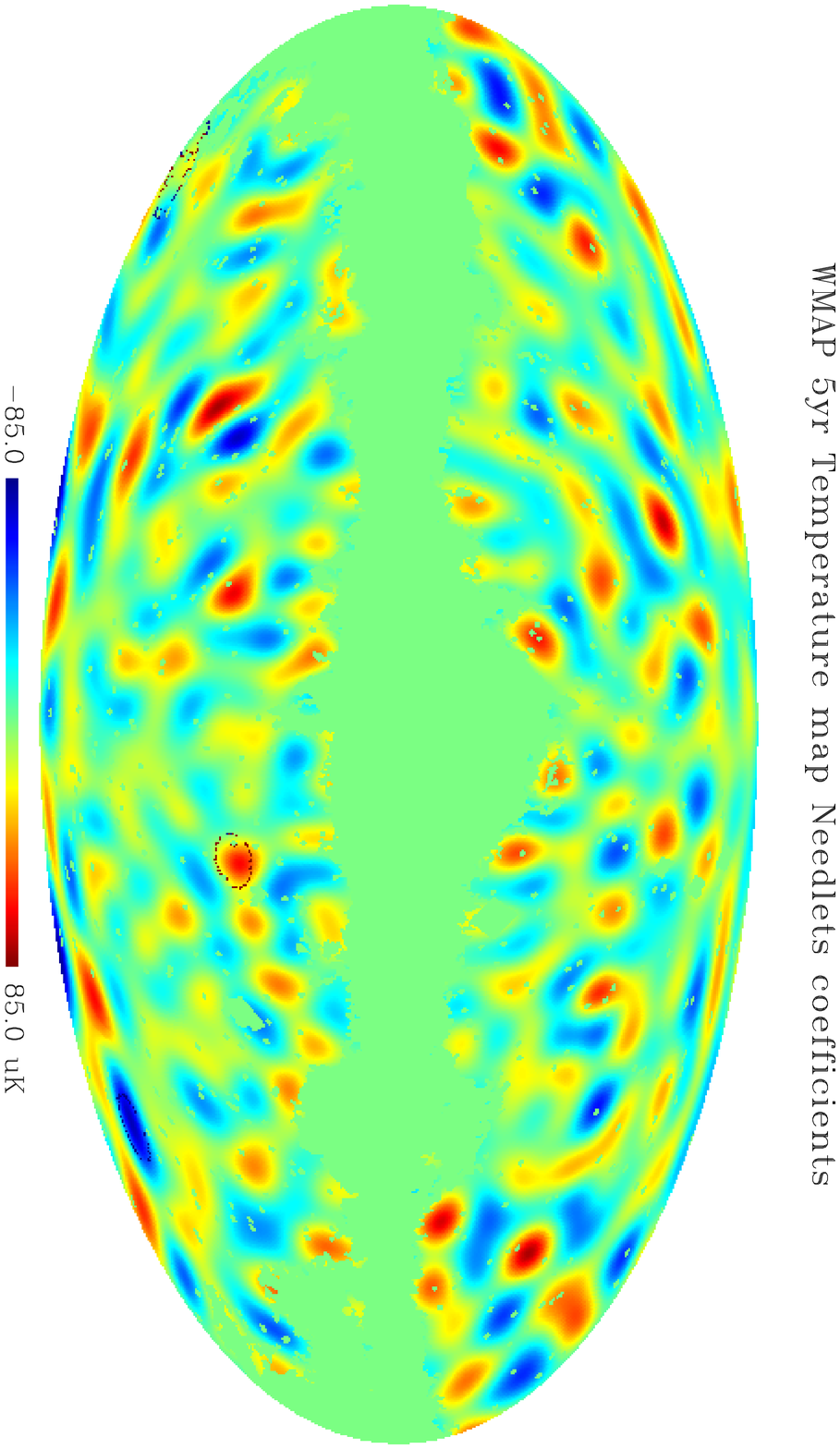}
\incgr[width=0.6\columnwidth, angle=90]{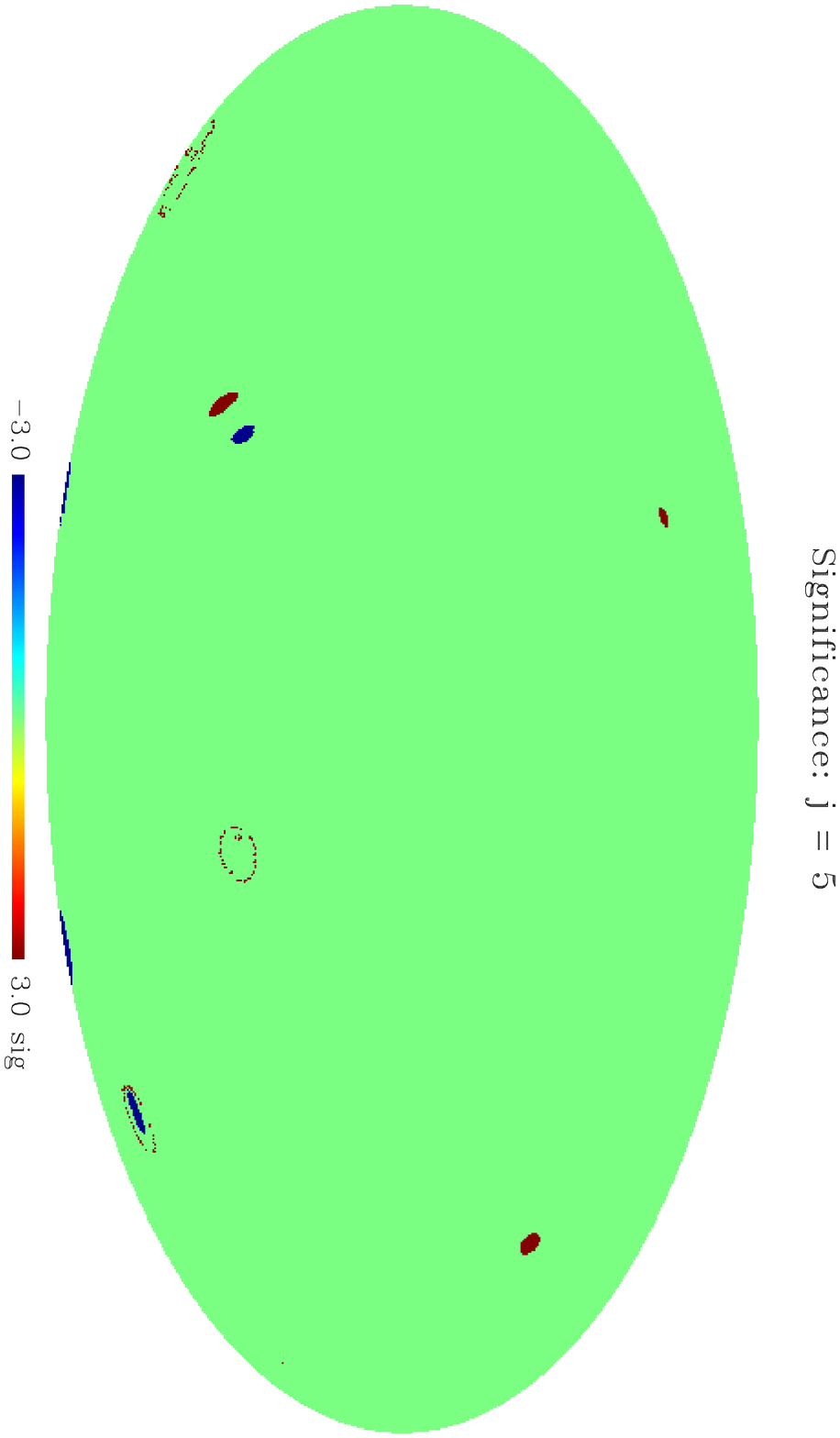}
\caption{Needlet coefficients of the WMAP 5-year CMB temperature map. On the left, from top $j=3$, $j=4$ and $j=5$ are plotted. The set of needlets is characterized by $B=1.8$. Each
 pixel represents the coefficient for the needlet function computed at $\xi_{jk}$, where $k$ identifies the pixel in the Healpix ordering.
The effect of the applied KQ$75$ mask is clearly visible, setting to zero the value of each pixel that belongs to the mask.
 It is interesting
to notice that needlet coefficients highlight the presence of the well-known cold spot in the southern hemisphere,
as well as a hot spot localized
in the southern hemisphere closer to the mask. Needlet coefficients for $j=5$ show the cold spot pretty well,
while the hot spot is weaker. Another couple of hot/cold spots appear above the Galactic plane.
On the right, from the top, significance of the needlet coefficients for $j=3$, $j=4$ and $j=5$.
The three maps show the significance $S_{jk}$ above the threshold of $3$.
This allows us to localize in $\ell$-space the contribution of the hot spot that results to be in the range
of multipoles between $\ell=6$ and $\ell=18$. Computing the coefficients for $j=6$ and observing that the cold spot,
if present, does not have a high statistical significance, we can deduce the range of multipoles covered by the cold spot:
between $\ell=6$ and $\ell=33$.}
\label{fig:need_coeff}
\end{figure*}

In Fig.~\ref{fig:need_coeff}, we represent the  needlet
coefficients for the cases of $j=3$ to $j=5$. These three
frequencies are particularly interesting since they probe the
low-multipole region, highlighting a peculiar pattern of
anisotropies in the southern hemisphere. When $j=4$, we recognize
two spots, one hotter and one colder than the average CMB
fluctuations. While the latter is a detection of the so-called
\emph{cold spot}, a feature studied in detail in the literature as a
source of non-Gaussianity in the CMB map
\cite{Vielva2004NG,Larson2004HCspots,Cruz2005ColdSpot,Vielva2007ColdSpot,Cruz2007ColdSpotW3,Cruz2008,SmithHuterer2008},
the hot spot has not had the same scrutiny with only a minor
description in Refs.~\cite{Naselsky2007,Vielva2007ColdSpot}.

In fact, the $j=3$ and $j=5$ needlet coefficients represent a
cross-validation of the existing literature, as they also highlight
a second hot spot centered at $(g_l=173,g_b=-46)$ (as pointed out in
Ref.~\cite{Naselsky2007}) and a minor cold spot centered at
$(g_l=80,g_b=-33)$ (observed in Ref. \cite{Vielva2007ColdSpot}), that actually appears with an adjacent hot spot.
In Fig.~\ref{fig:spots} the three spots
we used in the angular power spectrum analysis are shown. We stress
that our identification follows uniform criteria with the same
technique, quite differently from some of the existing
literature. Moreover, as explained above we are also able to
identify exactly the range of multipoles where these features
contribute to the angular power spectrum of CMB anisotropies, due to
the specific needlet properties.

\bfg[htb]
\incgr[width=.5\columnwidth, angle=90]{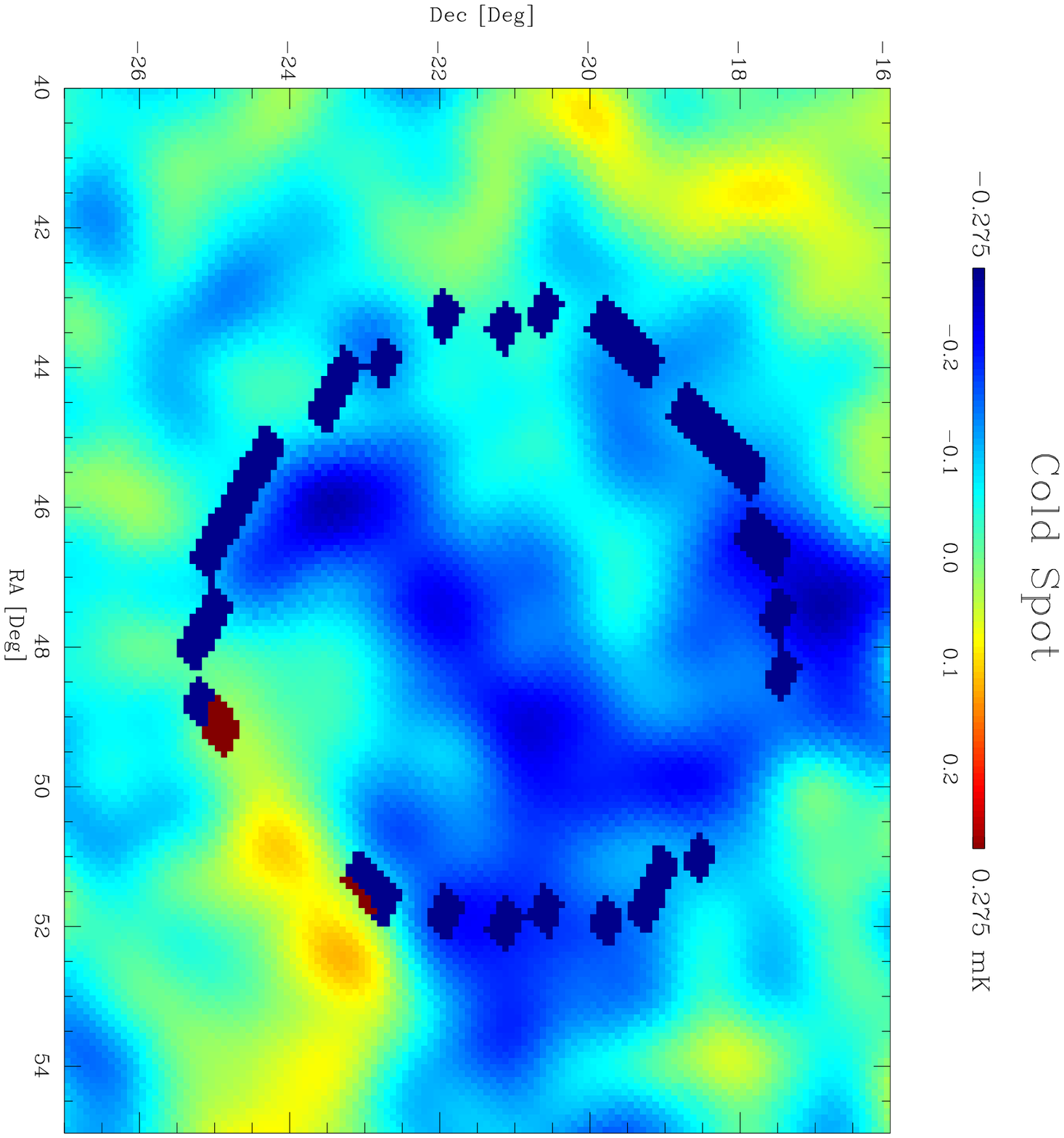}
\incgr[width=.5\columnwidth]{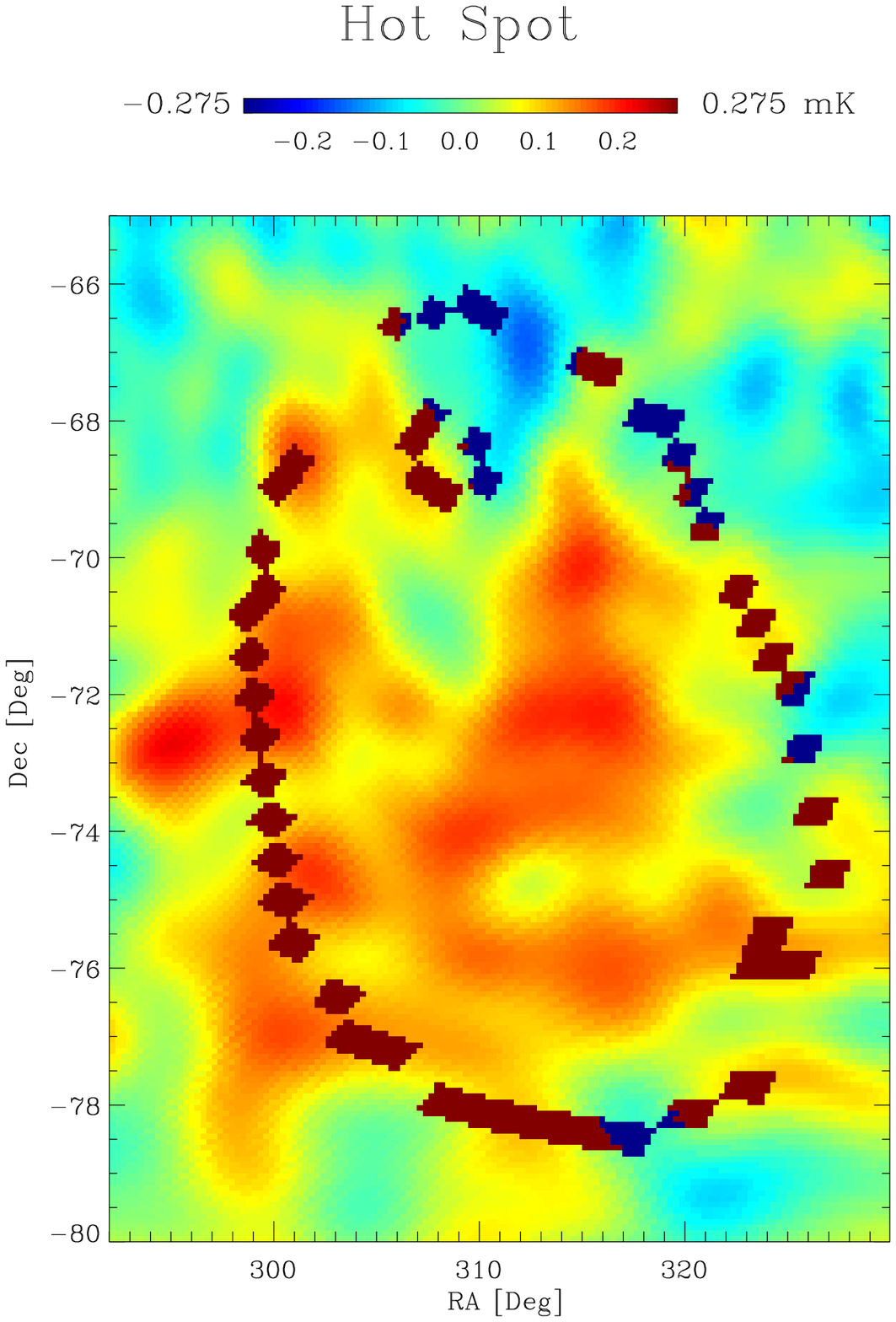}
\incgr[width=.5\columnwidth]{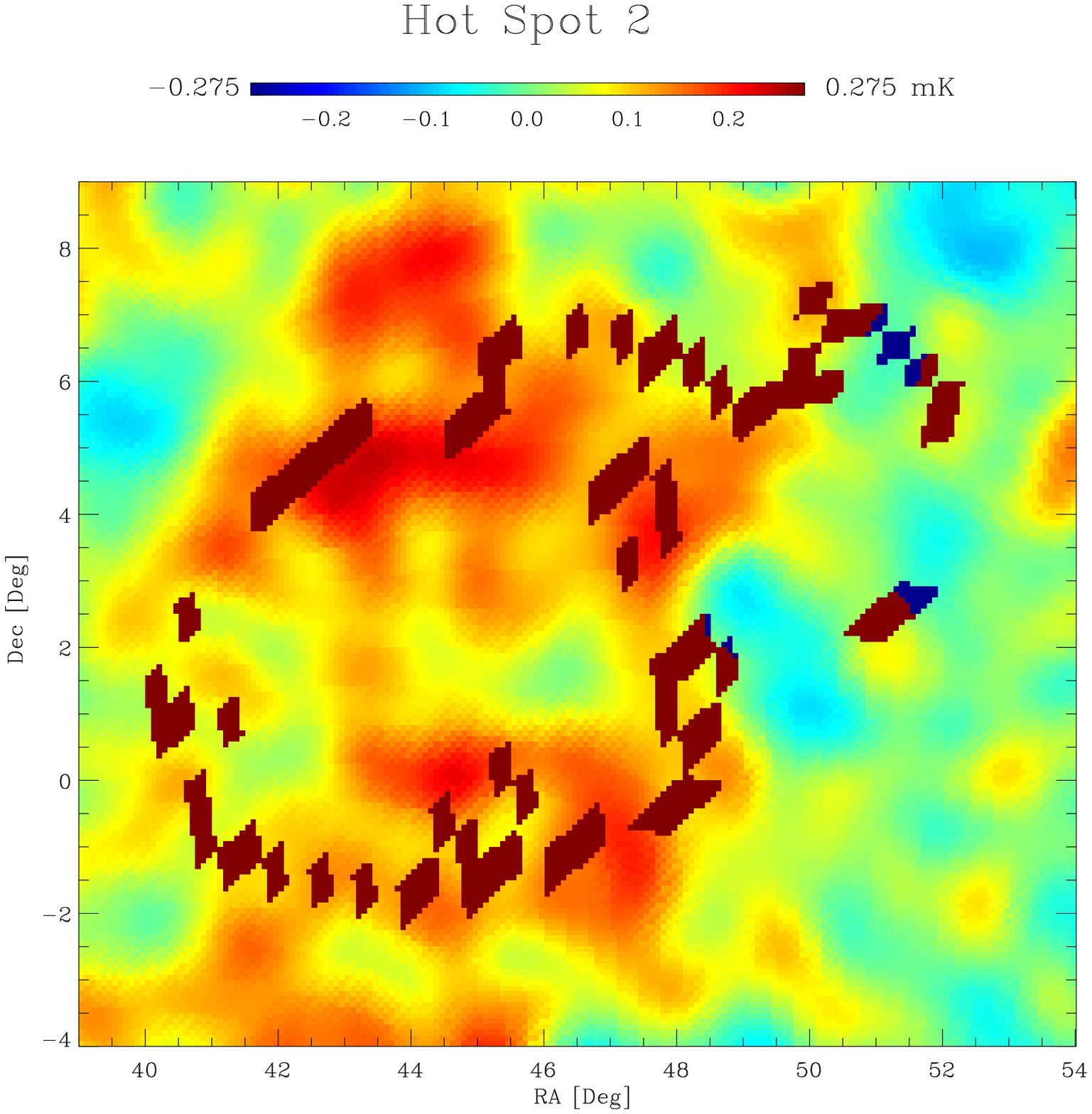}
\caption{Detail of the spots. From the top, the cold spot at $(g_l=209,g_b=-57)$, the hot spot at $(g_l=320,g_b=-34)$ present at $j=4$, and the second hot one at $(g_l=173,g_b=-46)$ measure at $j=3$. The true colors are altered by the use of the masks we employed in the analysis to highlight the region above three sigma level.}
\label{fig:spots}
\efg

To establish the significance of the features we detected, we
consider a Monte Carlo analysis by performing a large set of simulations (1000)
 of a Gaussian CMB sky with the same angular
power spectrum as the WMAP 5-year best-fit cosmological parameters \cite{WMAPKomatsu2008};
we then compute the average and the standard deviation of the
distribution for each needlet coefficient. The expected distribution is Gaussian with zero mean,
(see \cite{Baldi2006}), in good agreement with our simulations. We focus on the statistic
\be
  S_{jk}=\frac{|\beta_{jk}-<\beta_{jk}>|}{\sigma_{\beta_{jk}}},
\ee
where $\sigma_{\beta_{jk}}$ is the usual standard deviation of the distribution.
In Fig.~\ref{fig:need_coeff}, we show that the
hot and cold spots exceed three sigma level at $j=4$. The cold spot
appears significantly both in the needlet coefficients for $j=4$ and
$j=5$, and thus its impact can be reckoned to span the range between
$\ell=6$ and $\ell=33$ (Table~1). In Table \ref{tab:spots_sign} the significance values of the three anomalous spots are quoted.

\btb[h]
\begin{center}
\begin{tabular}{|c|c|c|c|}
\hline
{\bf Property / Spots} & {\bf Cold Spot} & {\bf Hot Spot}  & {\bf Hot Spot} \\
\hline
$(g_l,g_b)$      &$(209,-57)$ & $(320,-34)$ & $(173,-46)$ \\
\hline
$j$              & 4 & 4 & 3 \\
\hline
$S_{jk}^{max}$ & $(-)3.72$ & $3.56$ & $3.24$ \\
\hline
\end{tabular}
\caption{Main properties of the spots highlighted in our analysis.}
\label{tab:spots_sign}
\end{center}
\etb

It may be suspected that the hot spot we located could be a spurious effect due to oscillations in the needlet function.
We considered this explanation, but we concluded that the distance at which the hot and cold spot appear is
greater than the needlet oscillation range. We aim at a further investigation of this issue in our ongoing work.

\begin{figure*}[thb]
\incgr[width=0.6\columnwidth, angle=90]{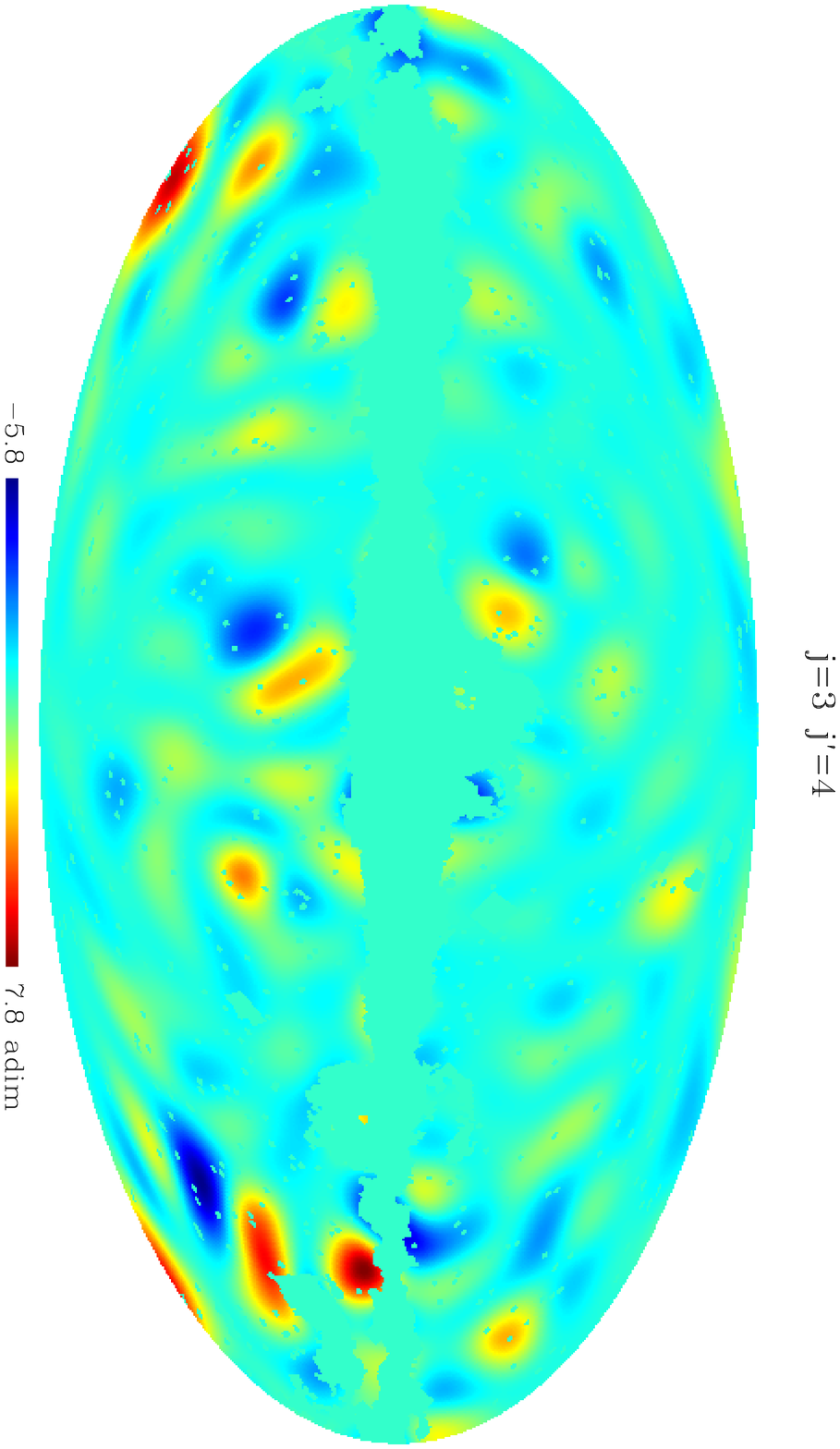}
\incgr[width=0.6\columnwidth, angle=90]{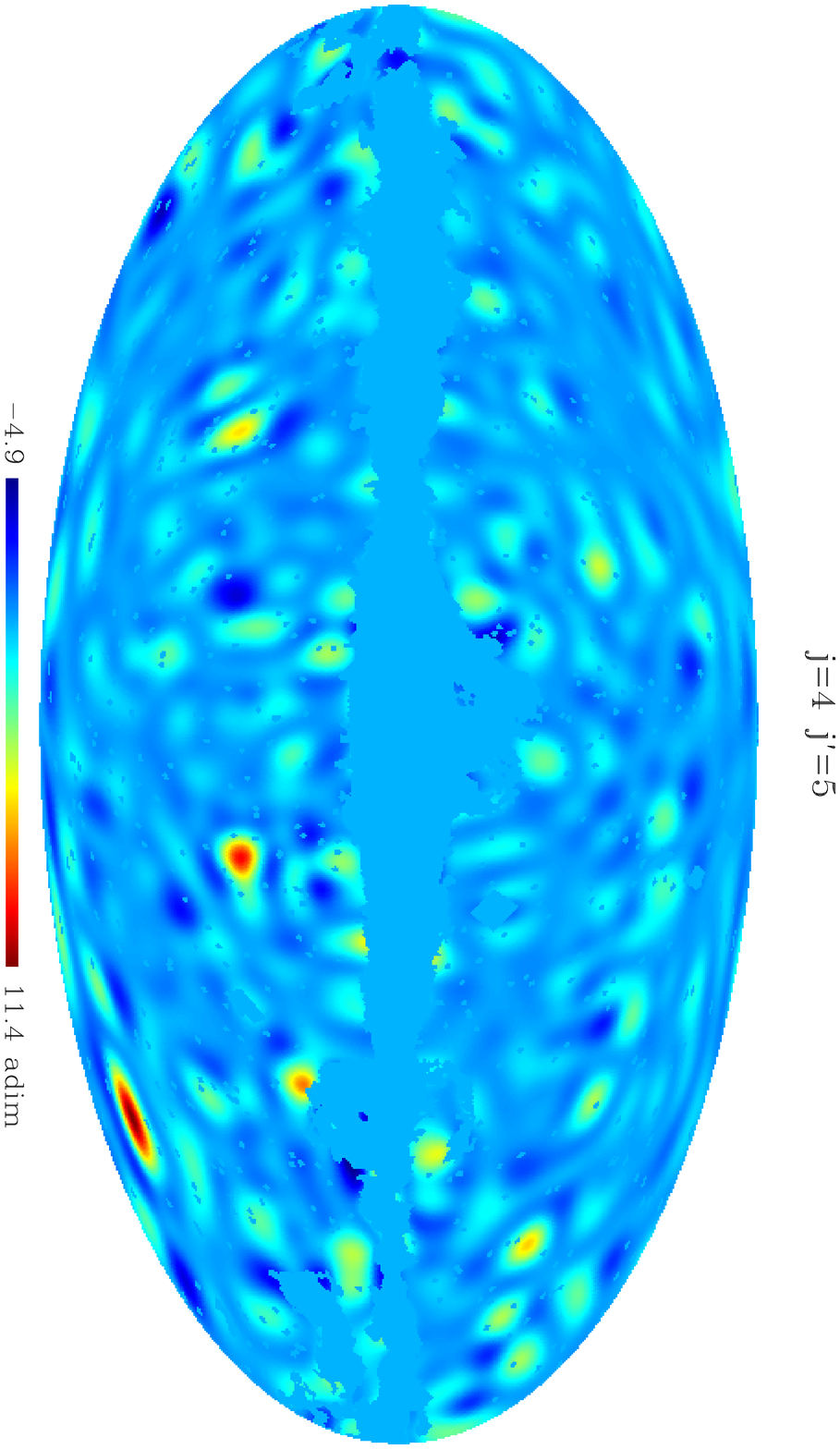}
\incgr[width=0.6\columnwidth, angle=90]{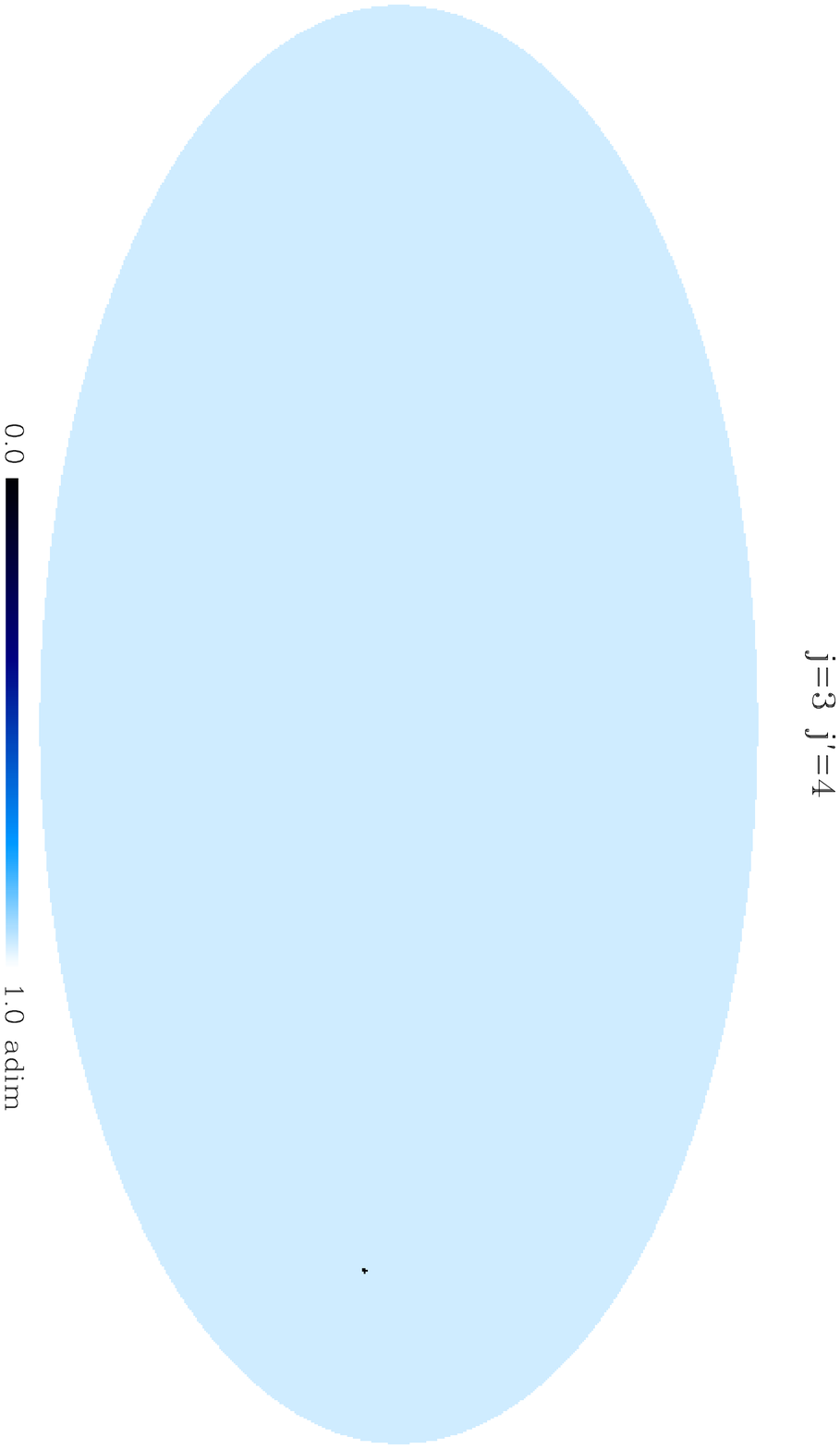}
\incgr[width=0.6\columnwidth, angle=90]{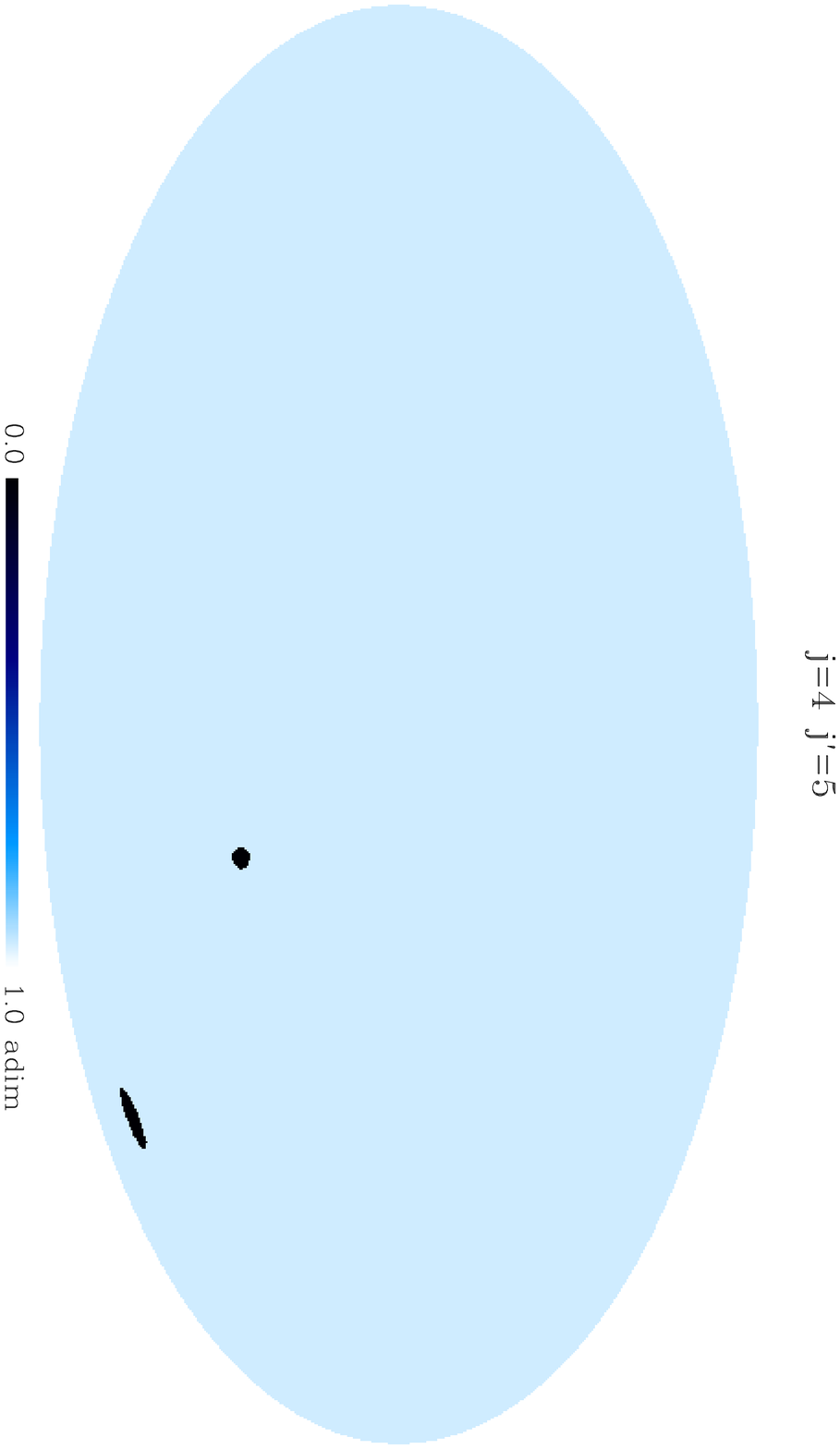}
\caption{$\Gamma$ statistics for $j=3$ and $j^\prime=4$ (left-hand side) and with $j=4$ and $j^\prime=5$, (right-hand side). Both the cold spot and the hot spot are clearly visible in the maps, but when a threshold of $\Gamma_k^{jj^\prime}>7.5$ is applied their signal is strong only on the $j=4$ and $j^\prime=5$ maps.}
\label{fig:need_corr}
\end{figure*}

As a test for the joint significance of needlet coefficients, we consider the statistic
 \be
  \label{eq:corr_sig}
  \Gamma_k^{jj^\prime} = \frac{\beta_{jk}\,\beta_{j^\prime k} - <\beta_{jk}\,\beta_{j^\prime k}>}{\Sigma_{k}^{jj^\prime}}
\ee
where $\Sigma_{k}^{jj^\prime}$ is the second moment of the
distribution. Of course, the statistical distribution of $\Gamma$
cannot be taken as Gaussian. Hence, to analyze its statistical
significance we used again a set of Monte Carlo
simulations and we followed the procedure described in Sec.~\ref{sec:gauss_check}.
Figure \ref{fig:need_corr} shows the
values of these statistics for the pair $j=3$ and $j=4$, left panel, and for $j=4$
and $j=5$, right panel. Results are expressed directly in terms of the
statistic defined in Eq.~\ref{eq:corr_sig} in the upper panels;
in the lower panels a threshold of $7.5$ for $\Gamma_k^{jj^\prime}$
is adopted to underline the fact that the hot and cold spots identified by the mask $j=4$ are the
most significant.

%\bfg[thb]
%\efg

% *****************************************
\subsection*{III.B  North-South Asymmetry}
\label{sec:need_corr}
\vspace{2cm}

We now wish to investigate the extent in which masking the
previously found features affects the asymmetry between the northern
and the southern hemisphere of the CMB sky. To see this, we recall that in
Refs.~\cite{Baldi2006,Pietrobon2006,Marinucci2008} it is shown that
\begin{displaymath}
\big\la\sum_k\beta_{jk}^2\big\ra=\sum_\ell\frac{(2\ell+1)}{4\pi}b^2\Big(\frac{\ell}{B^j}\Big)\mathcal{C}_\ell
\end{displaymath}
whence $\beta_j\equiv\sum_k\beta_{jk}^2$ is an unbiased estimator for the weighted angular power spectrum. In
\cite{Baldi2006}, Sec.~7, it is also shown that this statistic is
approximately Gaussian (after centering and normalization) at high
frequencies, see also \cite{Delabrouille2008maps,Fay2008b} for
extensions and related work. Developing this idea, we computed here
the total power in each hemisphere by taking the sum of squares of
the needlet coefficients at $j=3$ and $j=4$ extracted from the
masked ILC temperature map
\be
  \label{eq:34power}
  C_{3,4} = \sum_{j=3}^4\sum_k \beta_{jk}^2 = \sum_{j=3}^4\beta_j.
\ee
We label the mask associated with the $j=3$ and $j=4$
features we found as ``$j3j4$'', and we include it in addition to
standard WMAP masks.

The difference is measured by computing
\be D \equiv \frac{1}{V}\left(C_{3,4}^S-C_{3,4}^N\right)\,
,
  \label{eq:diffSN}
\ee
where $C_{3,4}^i$, $i=S,N$ is the quantity in Eq.~\ref{eq:34power},
where the needlet coefficients are restricted to either the northern or the southern hemisphere (in the
Galactic coordinate system), normalized to the variance $V$ of the whole sky. The latter actually turn out to be the cosmic variance of the CMB power spectrum binned with $b_{\ell,34}^2$. In some sense, D is measuring the difference
between two local versions of the power spectrum estimator; such local estimators can indeed be rigorously justified,
in view of the uncorrelation properties of needlets in pixel space (see \cite{Baldi2007}).
In practice, we defined the pixels in
the northern hemisphere as those outside the mask characterized by
$\theta < \pi/2$ and pixels in the southern hemisphere as those outside the mask,
with $\theta > \pi/2$.

In Fig.~\ref{fig:betaj} we plot $\beta_j$ as defined in Eq.~\ref{eq:34power} extracted from the whole sky as well as those measured in each hemisphere, both including the spots and masking them. The southern hemisphere shows an excess of power compared to the northern one that is reduced by a factor of 2 when the ``j3j4'' mask is applied. Notice that, as expected, the power measured in the north region is not affected by the masking procedure. In the lower panel of Fig.~\ref{fig:betaj} we quantify this effect computing the difference $\beta_j^S - \beta_j^N$, normalized to the variance of the estimator.
\bfg
\incgr[width=\columnwidth]{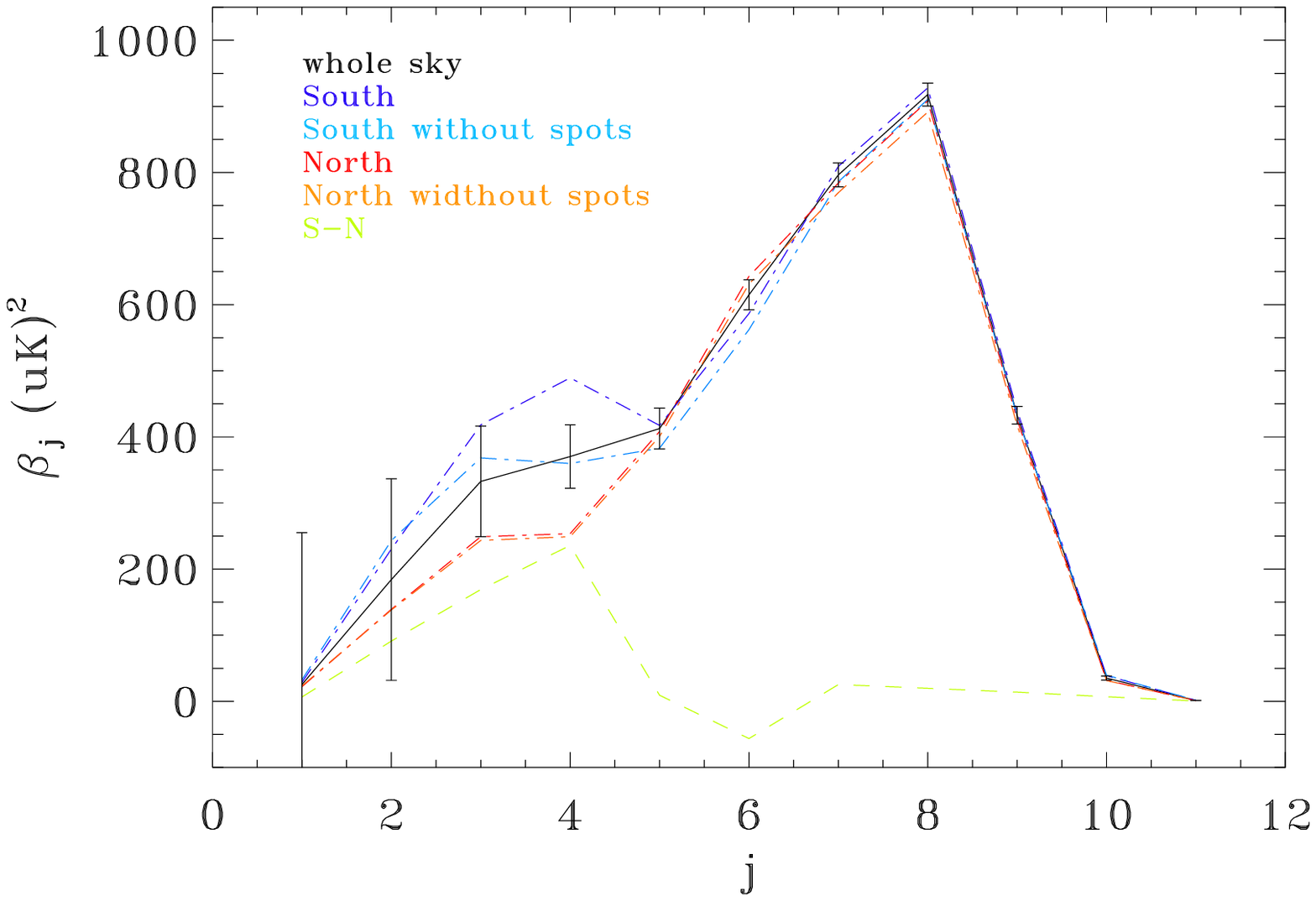}
\incgr[width=\columnwidth]{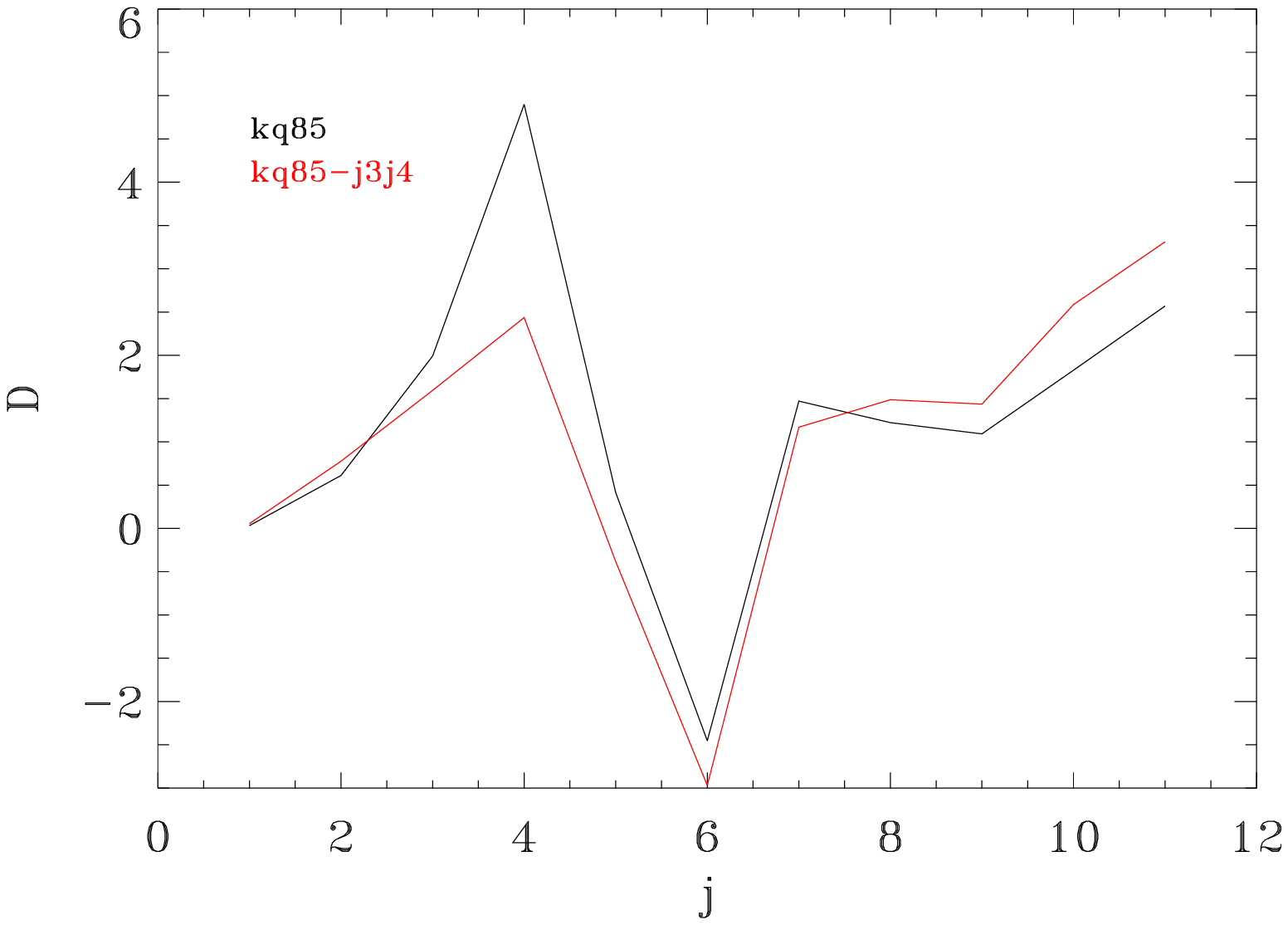}
\caption{In the upper panel $\beta_j$ extracted from WMAP data when different sky cuts are applied are shown. The black solid line shows the signal for the whole CMB sky (kq85 is applied). The blue and red dashed lines show how the power is split between the two hemispheres. When the cold and hot spots detected are masked the excess of power in the southern region is decreased (light blue and orange dot-dashed lines). In the lower panel the significance of the difference $D$ is plotted.}
\label{fig:betaj}
\efg

The results are summarized in Table \ref{tab:asym_sign}. Our findings
are sensitive to the chosen sky-cut: when the KQ$85$ mask is used,
the asymmetry, measured in terms of a difference in the variance of
power, decreases from $4.26$ without the $j3j4$ needlet mask to
$1.97$ with needlet mask.
When the more aggressive mask KQ$75$ is
applied to CMB data (together with the $j3j4$ mask), the difference
is larger and the global asymmetry is further reduced:
from $4.37$ to $2.0$. Note that the north-south power variance
difference with just the WMAP team's masks, KQ75 and KQ85, is rather
small (4.26 vs 4.37); this suggests that the hemispherical
asymmetry cannot be explained by simply extending the galactic plane
mask. With the $j3j4$ mask we introduced (which cuts roughly 0.5\%
of the sky), we find a significant reduction of a factor of 2.

Given that we mask a smaller area on the sky than the 10\%
difference between KQ85 and KQ75, it seems rather likely that the
hemispherical asymmetry can be credited to features in the southern
hemisphere. While we have localized (some of) these features, this
does not establish by itself whether the asymmetry is primordial or
associated with fluctuations in our local universe. There exist
already several extended studies on the nature of the cold spot
(\cite{Masina2008,Sakai2008NLstruct,Granett2008SupVoids,Cembranos2008ExtDim,Genova-Santos2008Sz,Cruz2008,Hansen2007Rev});
for further statistical studies we make publicly available the
$j3j4$ mask.

\btb[htb]
  \begin{center}
  \begin{tabular}{|c|c|c|c|c|}
    \hline
    {\bf Mask} & {\bf$C_{3,4}$} & {\bf $C_{3,4}^N$}  & {\bf $C_{3,4}^S$} & {\bf D} \\
    \hline
    {\bf kq$85$ }      & 766 & 556 & 974 & 4.26 \\
    \hline
    {\bf kq$85$+j3j4 } & 673 & 544 & 802  & 1.97  \\
    \hline
    { \bf kq$75$}      & 703 & 502 & 908  & 4.37 \\
    \hline
    {\bf kq$75$+j3j4}  & 655 & 492 & 728  & 2.0   \\
    \hline
  \end{tabular}
  \caption{We report the values of total power (Eq.~\ref{eq:34power}) carried by needlets at $j=3$ and $j=4$ extracted on the ILC map. Four cases, corresponding to the different masks we applied in this analysis, are shown. The last column reports the significance $D$ as defined in Eq.~\ref{eq:diffSN}. It is interesting to notice how masking the hot and cold spots reduces the asymmetry by a factor greater than 2, while the ``$j3j4$'' mask covers $0.5\%$ of the sky only.}
  \label{tab:asym_sign}
  \end{center}
\etb

\bfg[htb]
\incgr[width=.8\columnwidth]{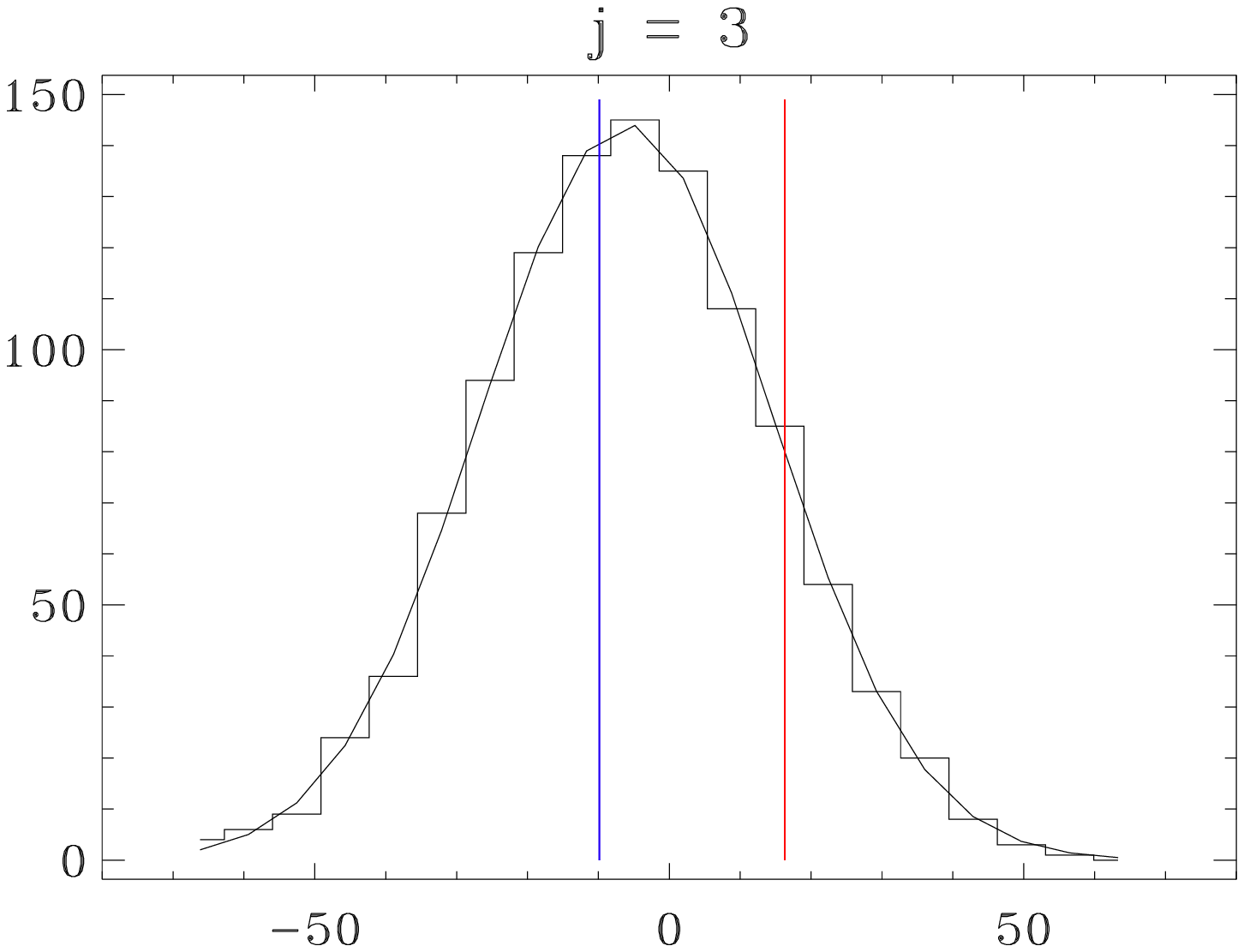}
\incgr[width=.8\columnwidth]{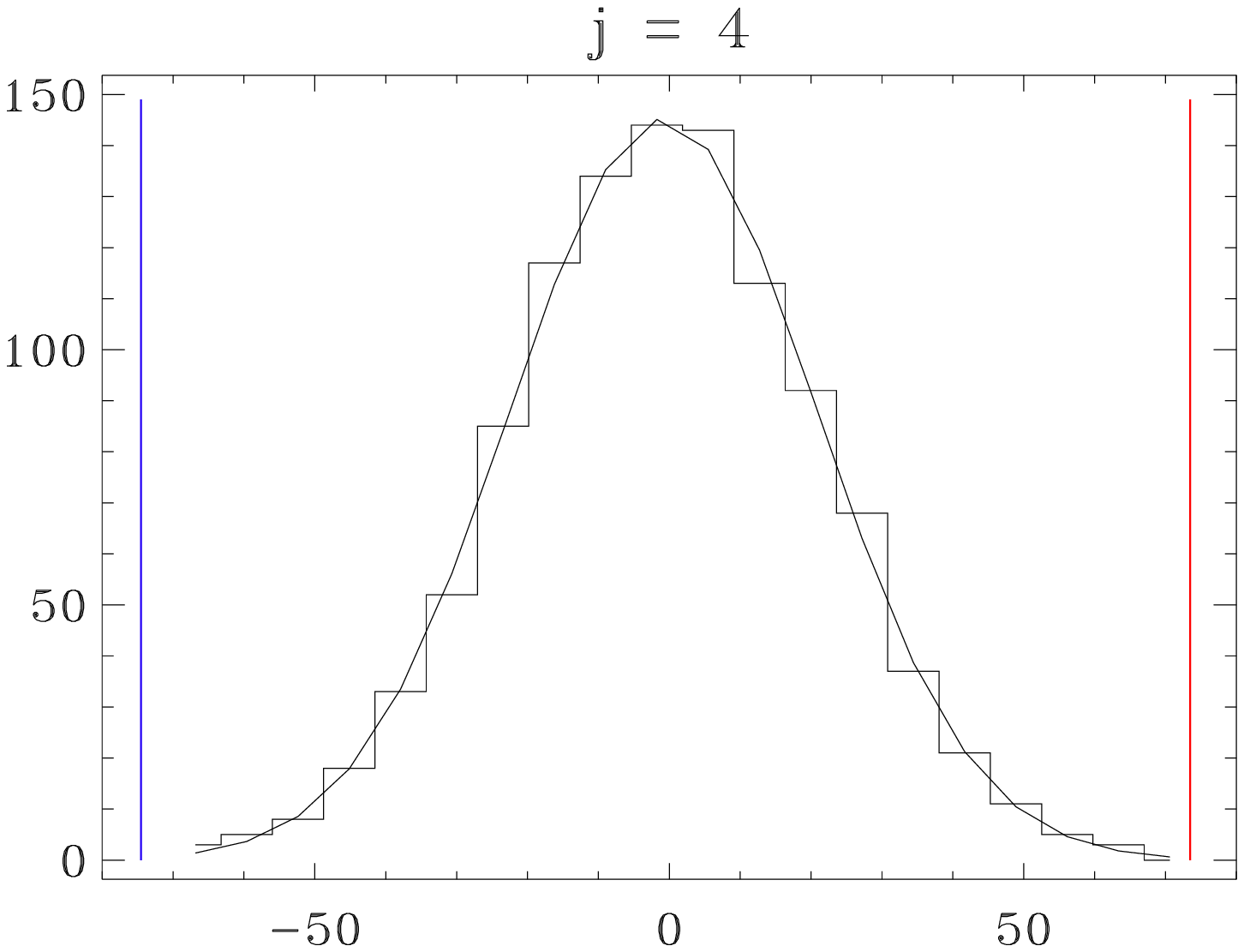}
\incgr[width=.8\columnwidth]{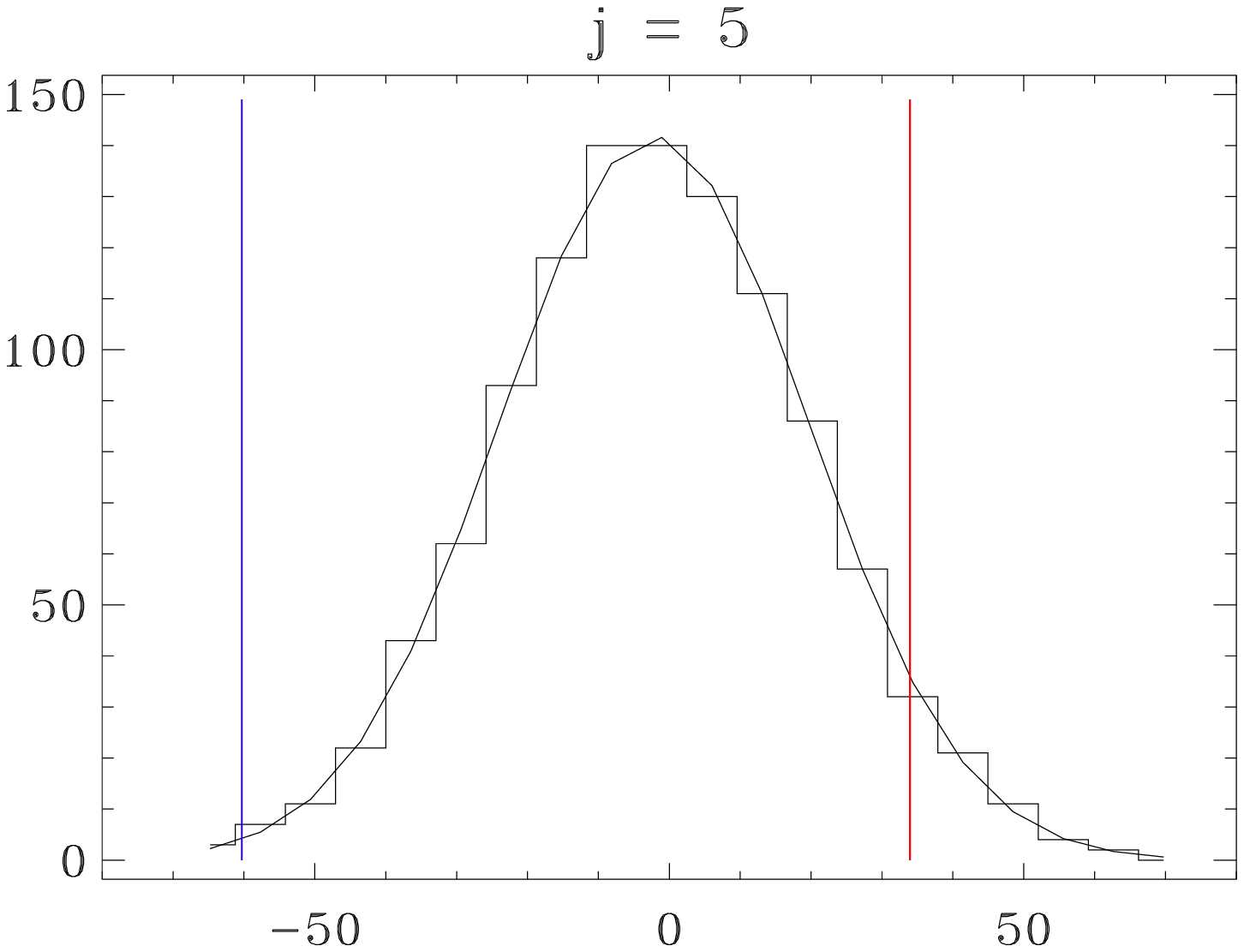}
\caption{From top distribution of the needlet coefficient for $j=3$, $j=4$ and $j=5$.  The blue (on the left-hand side) and red (on the right-hand side) vertical lines mark the values of the cold and hot spot respectively. They are both well beyond three sigma level for $j=4$. At $j=3$ he significance is basically zero, while at $j=5$ only the cold spot is still visible.}
\label{fig:gauss_check}
\efg

\subsection*{III.C  Further statistical analysis}
\label{sec:gauss_check}

As a further statistical cross-check,
we chose randomly needlet coefficients at different locations on the map and fit their distribution to the one derived from
1000 simulations where the KQ$85$ mask was applied.
The simulated results are in excellent agreement with a zero-mean Gaussian distribution,
as shown in Fig.~\ref{fig:gauss_check}. This result is of course expected, as the needlet coefficients
are a linear functional of the underlying temperature map. However, we report the figure as a further check to assure
that the procedure we followed to compute the significance of the spots is well justified.

We followed the same procedure also to quantify the significance of $\Gamma_k^{jj^\prime}$.
Of course, in this case simulations are indeed necessary,
because $\Gamma_k^{jj^\prime}$ is a nonlinear statistic and hence non-Gaussian.
In Fig.~\ref{fig:corr_dist} we provide some evidence on the significance
of the statistics we measured in the regions where the anomalous spots are located.
The curve is the fit to the distribution of needlet coefficients in simulated maps,
while the vertical lines mark the value measured in the WMAP temperature needlet coefficients.
 The simulated distribution has rather large tails; the statistical significance of
$\Gamma_k^{jj^\prime}$ at the locations corresponding to the spots is nevertheless rather high, with an estimated p-value of
$0.5\%$ for the cold spot and of $1.65\%$ for the hot spot.
We can still confirm that the signal is mostly peaked in the correlation between $j=4$ and $j=5$.

\bfg[htb]
\incgr[width=.8\columnwidth]{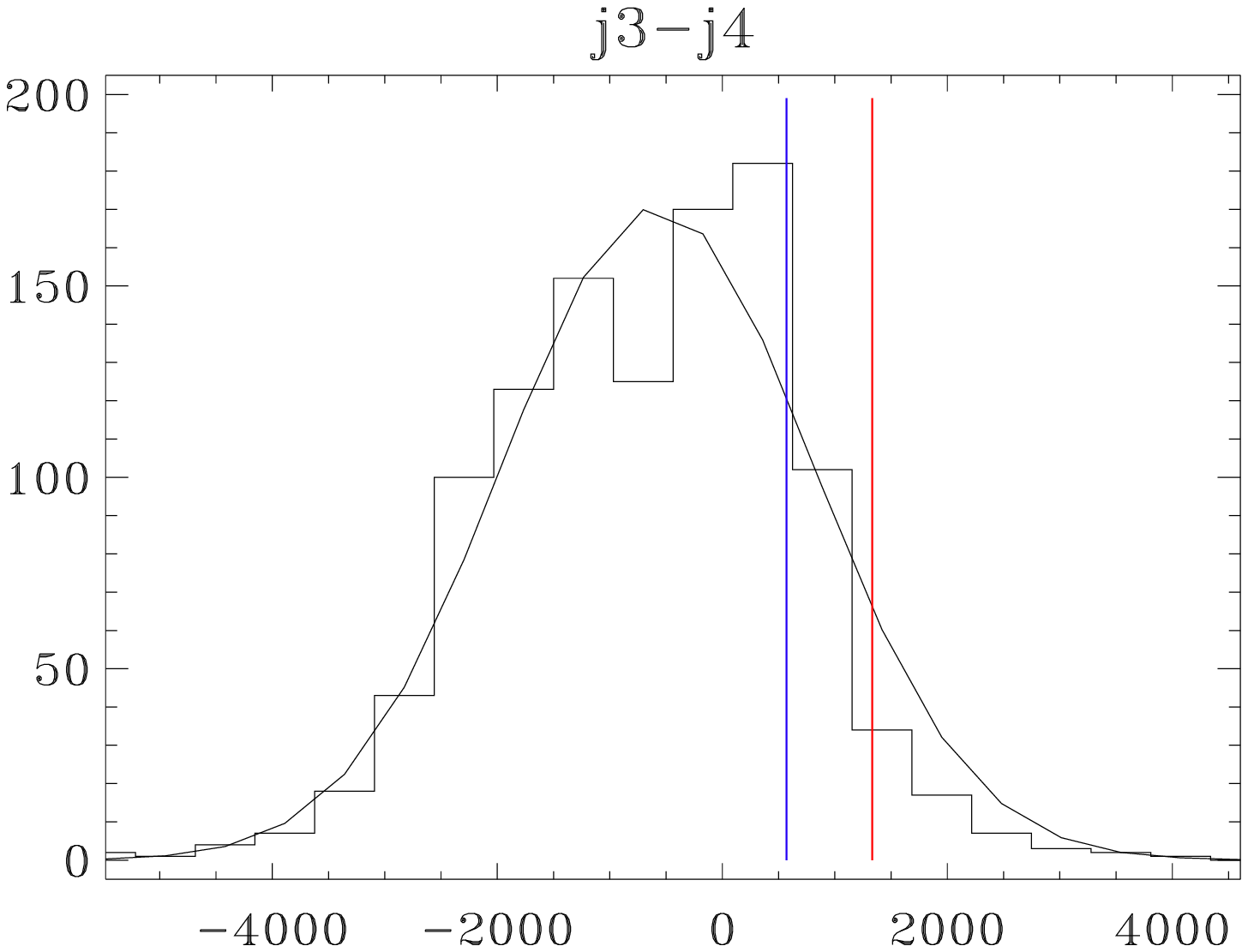}
\incgr[width=.8\columnwidth]{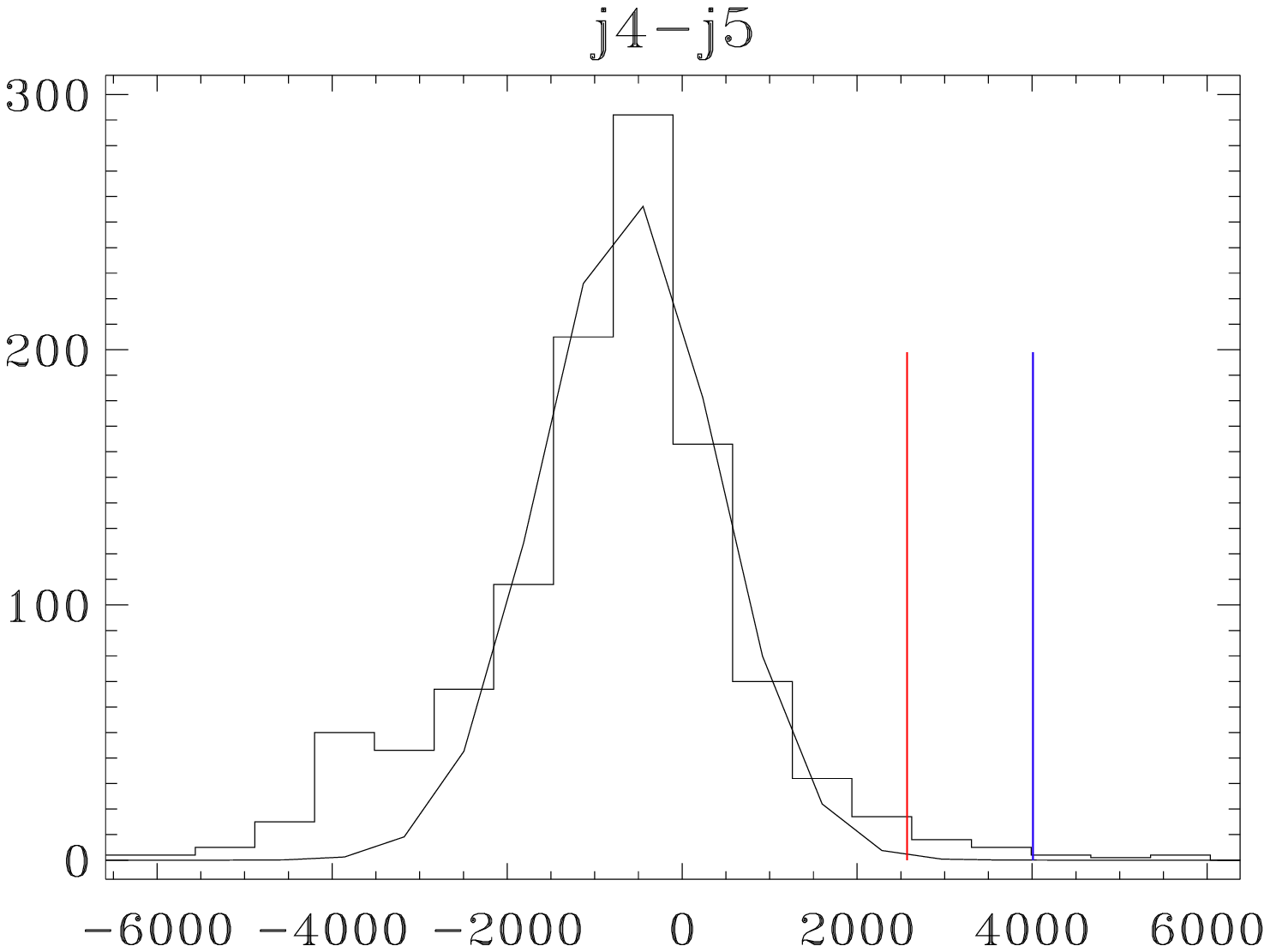}
\caption{Distribution of $\Gamma$ statistic for $(j,j^\prime)=(3,4)$ (top) and $(j,j^\prime)=(4,5)$. The two vertical lines mark the values of the hot and cold spot, the latter being the more significant. As expected, this is non-Gaussian, and is characterized by large non-Gaussian tails.}
\label{fig:corr_dist}
\efg

\bfg[htb]
\incgr[width=0.6\columnwidth, angle=90]{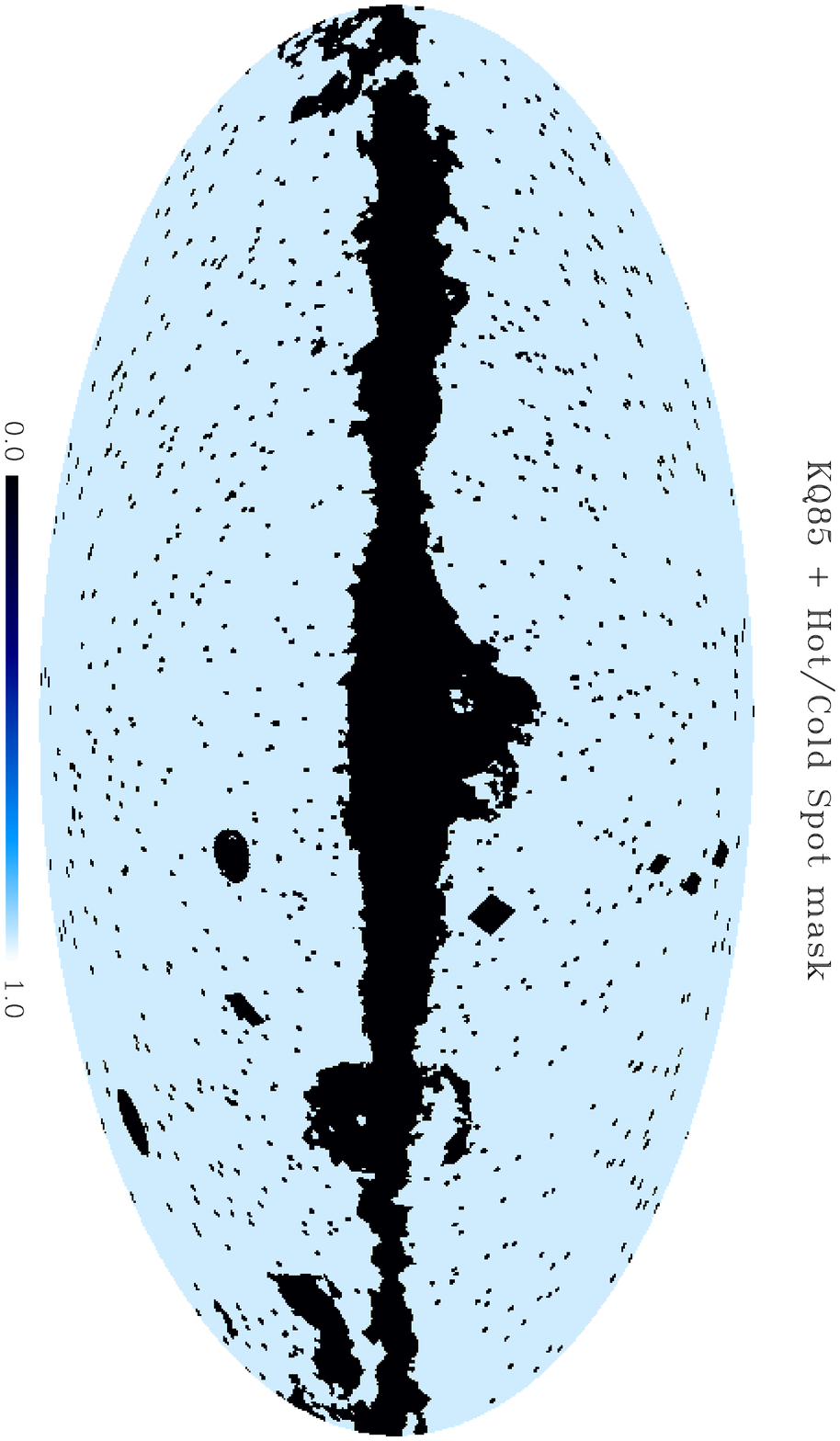}
\caption{Joint KQ$85$ and hot/cold spots mask applied to the WMAP ILC temperature map.}
\label{fig:KQ85HCS_mask}
\efg

\section{Impact on the CMB power spectrum}
\label{sec:cls}

The purpose of this section is to investigate the extent in which masking the hot and cold regions we found affects
the CMB power spectrum \cite{Masina2008}. To address this issue, we first estimate the angular power spectrum from
the ILC WMAP 5-year map after applying the KQ$85$ mask.
We then compare this result with the angular power spectrum resulting from a wider mask: the sum of KQ$85$ plus the
regions above three sigma level we discovered when performing our temperature analysis. We report in
Fig.~\ref{fig:KQ85HCS_mask} the resulting mask.

The effect of the different masking is not negligible, reaching the value of $12\%$ at low multipoles.
To check against systematics, we performed a Monte Carlo simulation of 200 CMB maps with the underlying
WMAP 5-year best-fit model \cite{WMAPKomatsu2008}, in order to estimate the mean effect of the joint mask.
We computed the average and the standard deviation to quantify the hot/cold spots effect.
The results are shown in Fig.~\ref{fig:cls_diff}.

\bfg[htb]
\begin{center}
\incgr[width=\columnwidth]{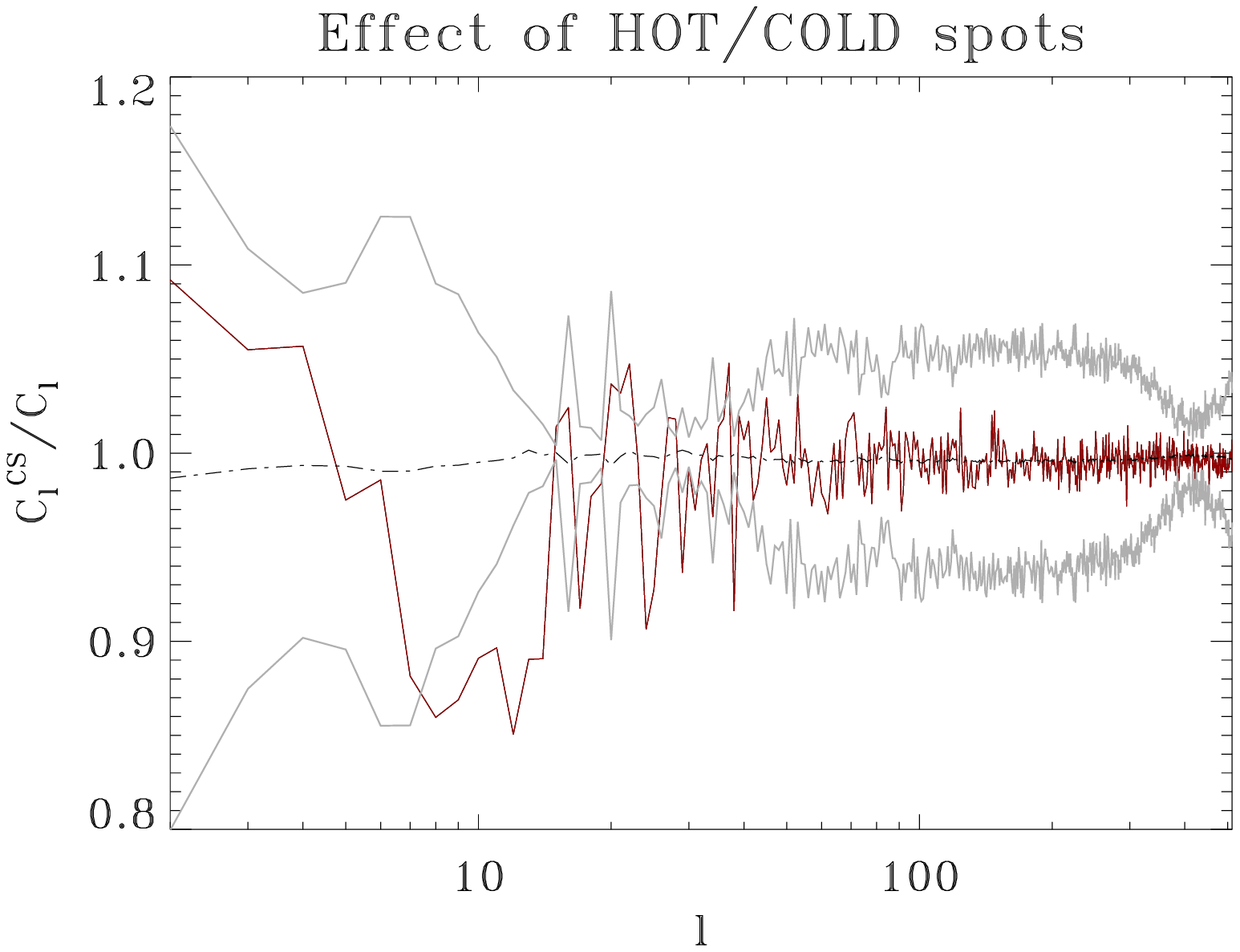}
\caption{Effect on the angular power spectrum due to the subtraction of the hot and cold spots in the CMB temperature map. The red solid line shows the difference in the $\mathcal{C}_\ell$; the gray dashed-dotted line represents the average modification of the simulation, while the solid lines mark the one sigma level. In the region between $\ell=8$ and $\ell=30$ the effect exceeds one sigma level: that region is the one where the effect of the cold spot is stronger.}
\label{fig:cls_diff}
\end{center}
\efg

Quite remarkably, the region where the signal is stronger is exactly the one where the cold spot is localized.
This may be interpreted as a confirmation of the localization properties of needlets in pixel  and harmonic spaces.
We believe the use of the ILC map is justified here, because  our signal peaks at low multipoles; however for completeness
we computed the same quantity from both
the W and V bands of WMAP 5 year, as well as using the map obtained
by \cite{deOliveiraCosta2008}\footnote{http://space.mit.edu/home/angelica/gsm/}. The result is fully consistent
with what we found using ILC, thus validating the procedure we followed.
The signal is shown in Fig.~\ref{fig:map_compar}

\bfg
\incgr[width=\columnwidth]{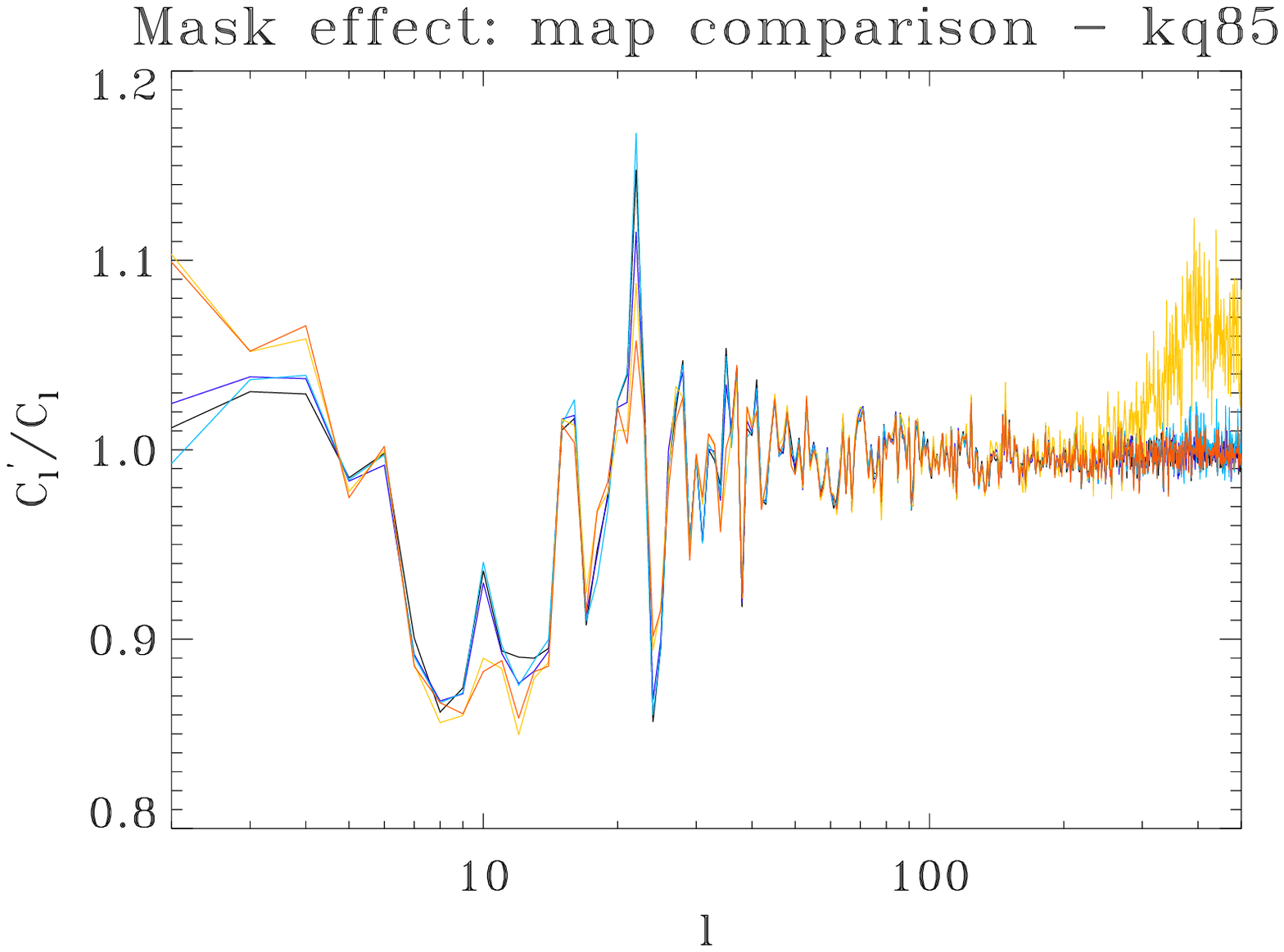}
\caption{Variation due to the combination of the hot and cold spots on five different power spectra: ILC the yellow curve, W the black curve, V the blue one, WV the light blue curve, and that extracted by the map reconstructed in \cite{deOliveiraCosta2008} in orange. Except for the WMAP ILC map at $\ell > 200$, the difference on the angular power spectrum
is consistent among a variety of maps. Beyond $\ell=200$ the ILC power spectrum shows features not compatible with other spectra, probably due to the way the different WMAP channels are combined.}
\label{fig:map_compar}
\efg

The next step has been to evaluate the effect that the  change in the angular power spectrum
induces on cosmological parameter estimates. Since the low multipoles region is affected by the largest variation,
we expect that changes might occur on the spectral index, $n_s$, the optical depth, $\tau$ and possibly on
the primordial fluctuation normalization amplitude $A_s$. Actually, since the variation is at maximum $12\%$
in an handful of multipoles, we expect a global variation on the power spectrum of roughly few parts on 1000.

The WMAP team performed the temperature analysis splitting the low multipoles and the high multipoles regions.
The former is probed by a Gibbs-sampling based Monte Carlo analysis \cite{Eriksen2004GibbsSampling}.
The high moments are instead investigated fitting the angular power spectrum extracted from the W and V bands.
To take into account the modifications due to the new masking we replaced the KQ$85$ mask used at low resolution with
the joint mask KQ$85$ plus the hot and cold spots (Fig.~\ref{fig:KQ85HCS_mask}), and multiplied
the angular power spectrum used for the analysis by the ratio between the $\mathcal{C}_\ell$ computed with the wider mask
and $\mathcal{C}_\ell$ obtained with the unmodified KQ$85$ mask applied to ILC,
in the range of multipoles $2-200$ (see Fig.~\ref{fig:map_compar}).

The results are shown in Table \ref{tab:cosm_pars}. In short, we do not observe a significant variation in any of
the cosmological parameters.

\begin{table*}[htb]
\begin{center}
\begin{tabular}{|c|c|c|c|}
\hline
Parameter        & WMAP5               & Hot/Cold spot masked ($j4$) & $j3$-$j4$ mask \\
\hline
$\mbf{\Omega_b h^2}$   & $\mbf{0.0227 \pm 0.0006}$      & $\mbf{0.0228 \pm 0.0006}$      & $\mbf{0.0228 \pm 0.0006}$ \\
\hline
$\mbf{\Omega_c h^2}$   & $\mbf{0.110  \pm 0.006}$       & $\mbf{0.109  \pm 0.006}$       & $\mbf{0.109  \pm 0.006}$  \\
\hline
$\theta_A$       & $1.040  \pm 0.003$       & $1.040  \pm 0.003$       & $1.040  \pm 0.003$  \\
\hline
$\mbf{\tau}$     & $\mbf{0.089 \pm 0.018}$  & $\mbf{0.091  \pm 0.017}$ & $\mbf{0.089  \pm 0.017}$  \\
\hline
$\mbf{n_s}$      & $\mbf{0.965 \pm 0.014}$  & $\mbf{0.966  \pm 0.014}$ & $\mbf{0.966  \pm 0.014}$  \\
\hline
$\mbf{ln(10^{10}A_s)}$ & $\mbf{3.18 \pm 0.05}$          & $\mbf{3.17   \pm 0.05}$        & $\mbf{3.17   \pm 0.05}$ \\
\hline
\end{tabular}
\caption{Effect of the wider mask on the $\Lambda$CDM six parameters. The difference due to the sum of KQ$85$ mask plus hot/cold spot mask is small.}
\label{tab:cosm_pars}
\end{center}
\end{table*}

\begin{figure*}[htb]
\incgr[width=.6\columnwidth, angle=90]{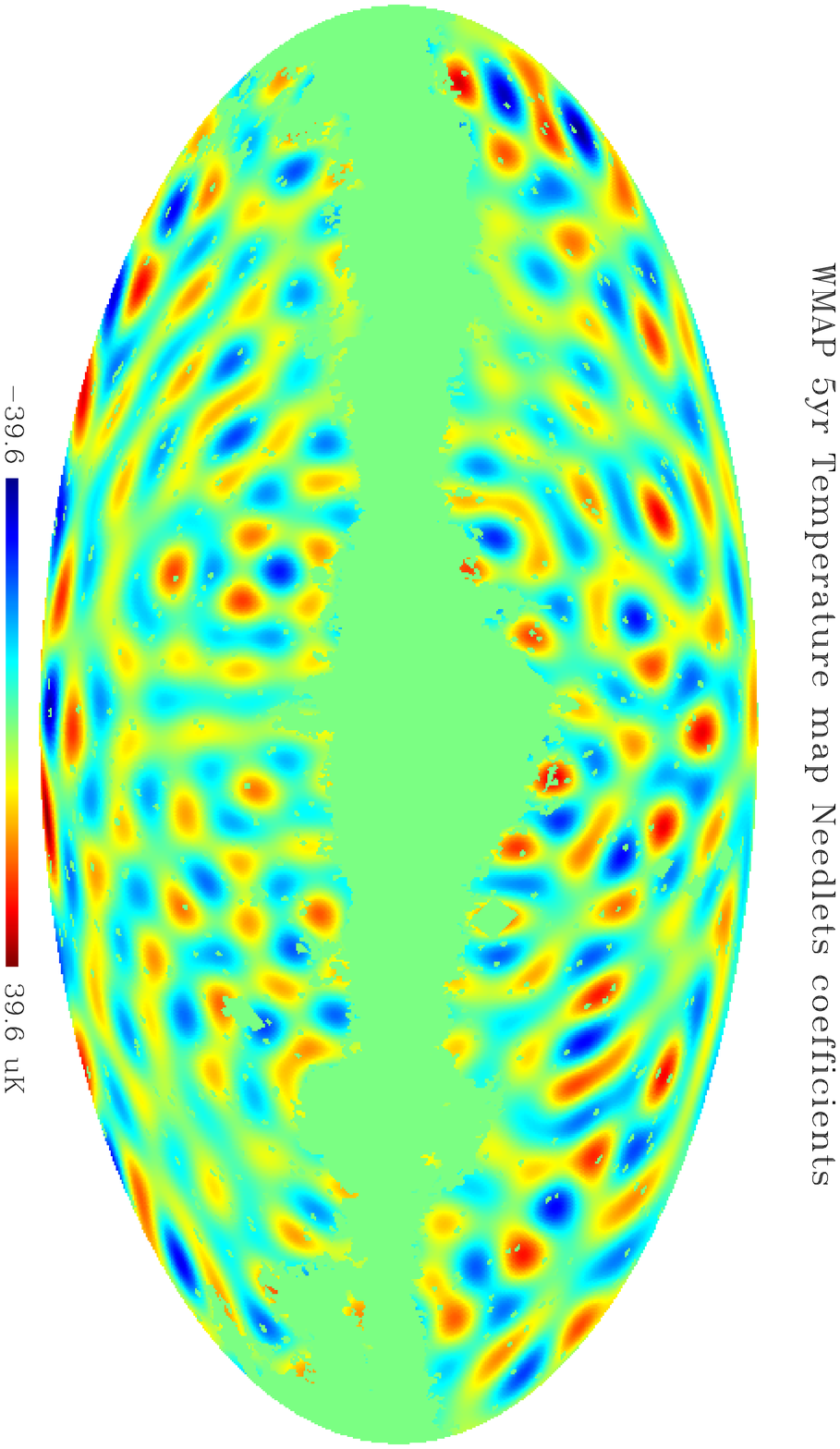}
\incgr[width=.6\columnwidth, angle=90]{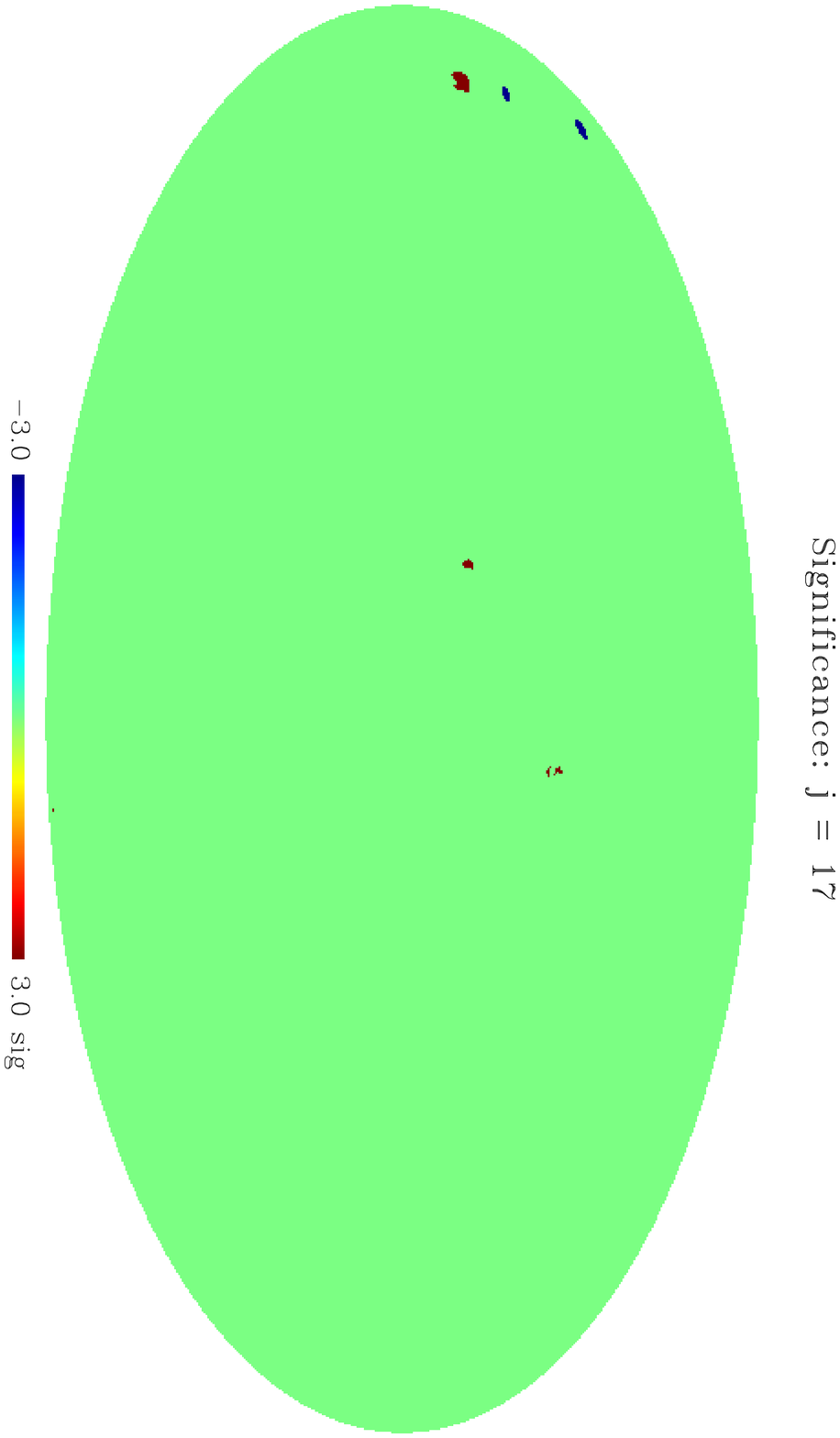}
\incgr[width=.6\columnwidth, angle=90]{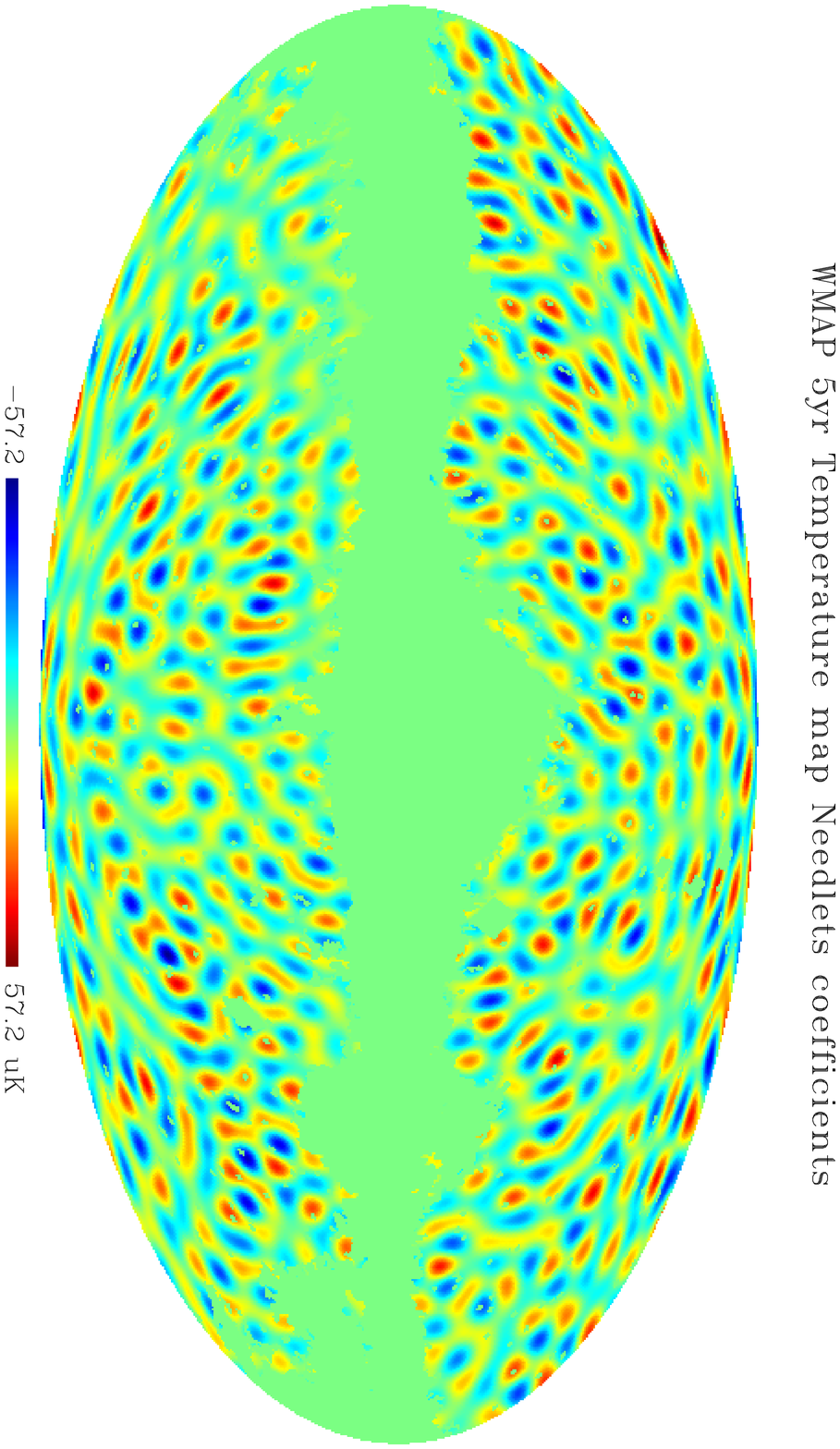}
\incgr[width=.6\columnwidth, angle=90]{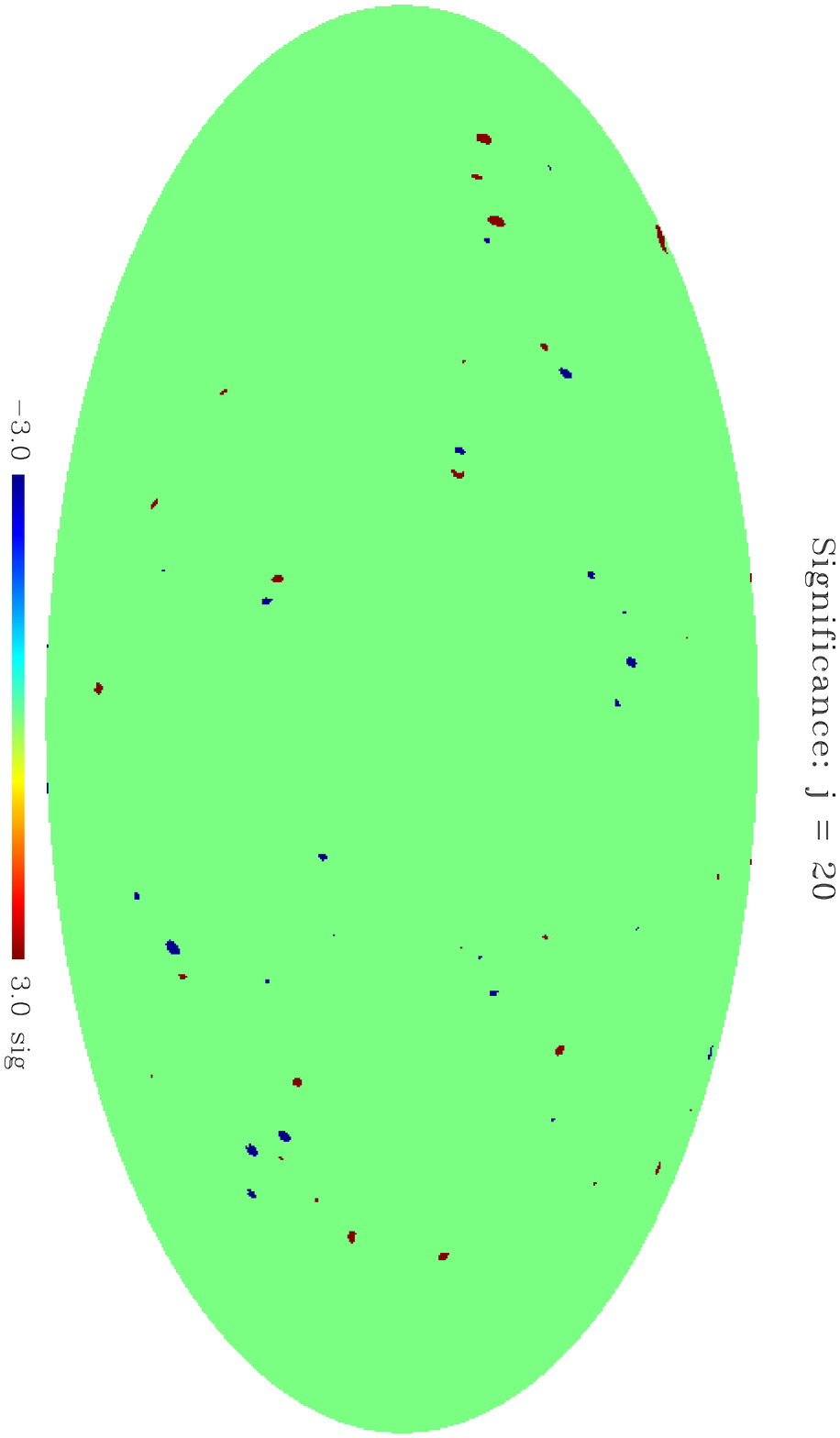}
\caption{Needlet coefficients and their significance at $j=17$ and $j=20$ for $B=1.2$.}
\label{fig:j17j20}
\end{figure*}

It is well known that the angular power spectrum measured by the WMAP Collaboration shows some interesting features at
low multipoles; in particular the range between $\ell=20$ and $\ell=24$ has a deficit in power with respect to the
prediction of the best-fit $\Lambda$CDM theoretical model, while that between $\ell=37$ to $\ell=44$ shows excess
power. To investigate these issues, we chose needlets corresponding to frequencies that match those
two intervals and we looked for coefficients exceeding the threshold of three sigma. More precisely, we select
 $B=1.2$ and we take $j=17$, $j=20$ to span the relevant ranges of multipoles.
Figure \ref{fig:b1.2profile} (upper panel) and Fig.~\ref{fig:j17j20}
show (respectively) the $b_\ell$ profile employed for this purpose and the corresponding needlet coefficients.

\bfg[htb]
\incgr[width=\columnwidth]{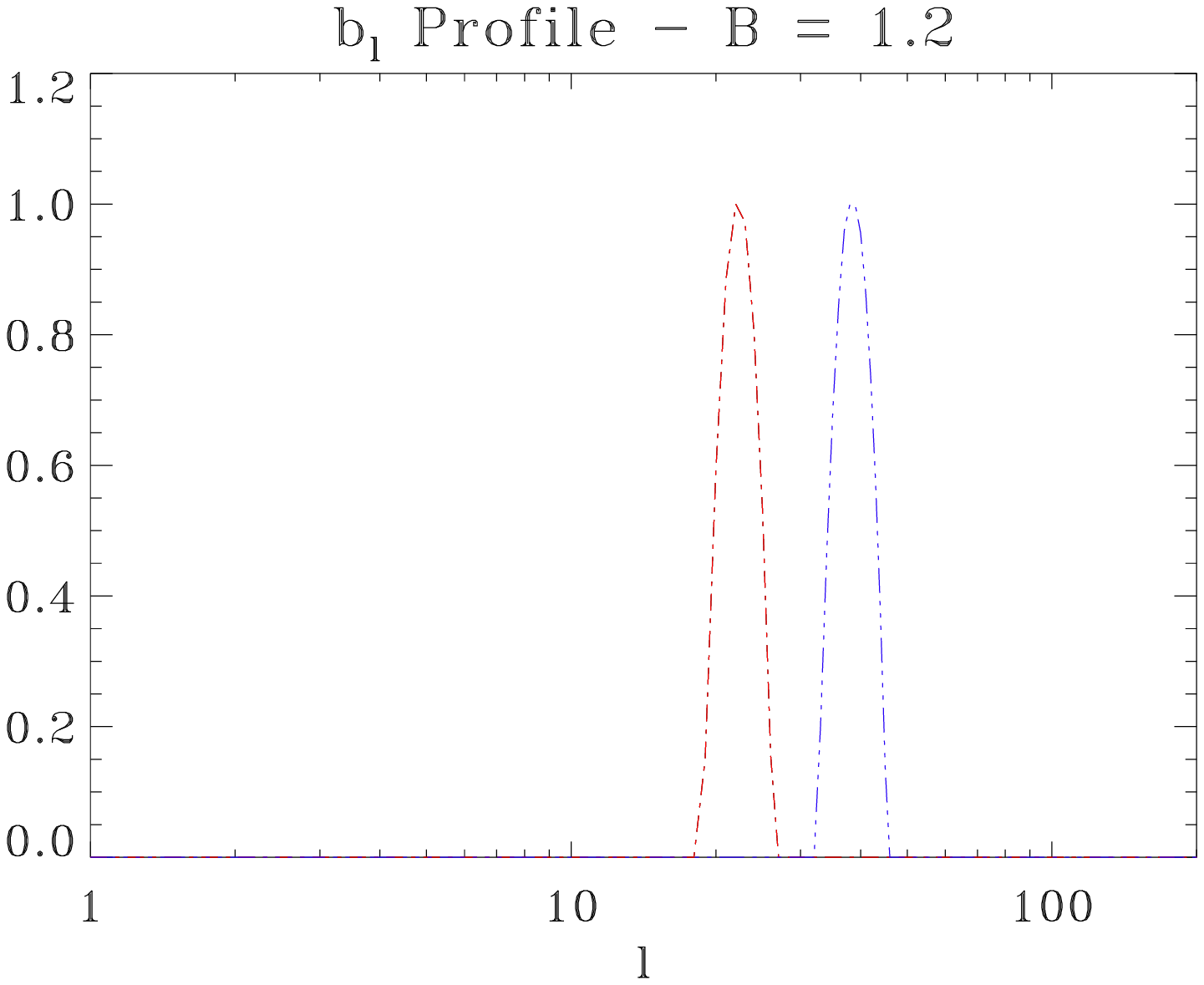}
\incgr[width=\columnwidth]{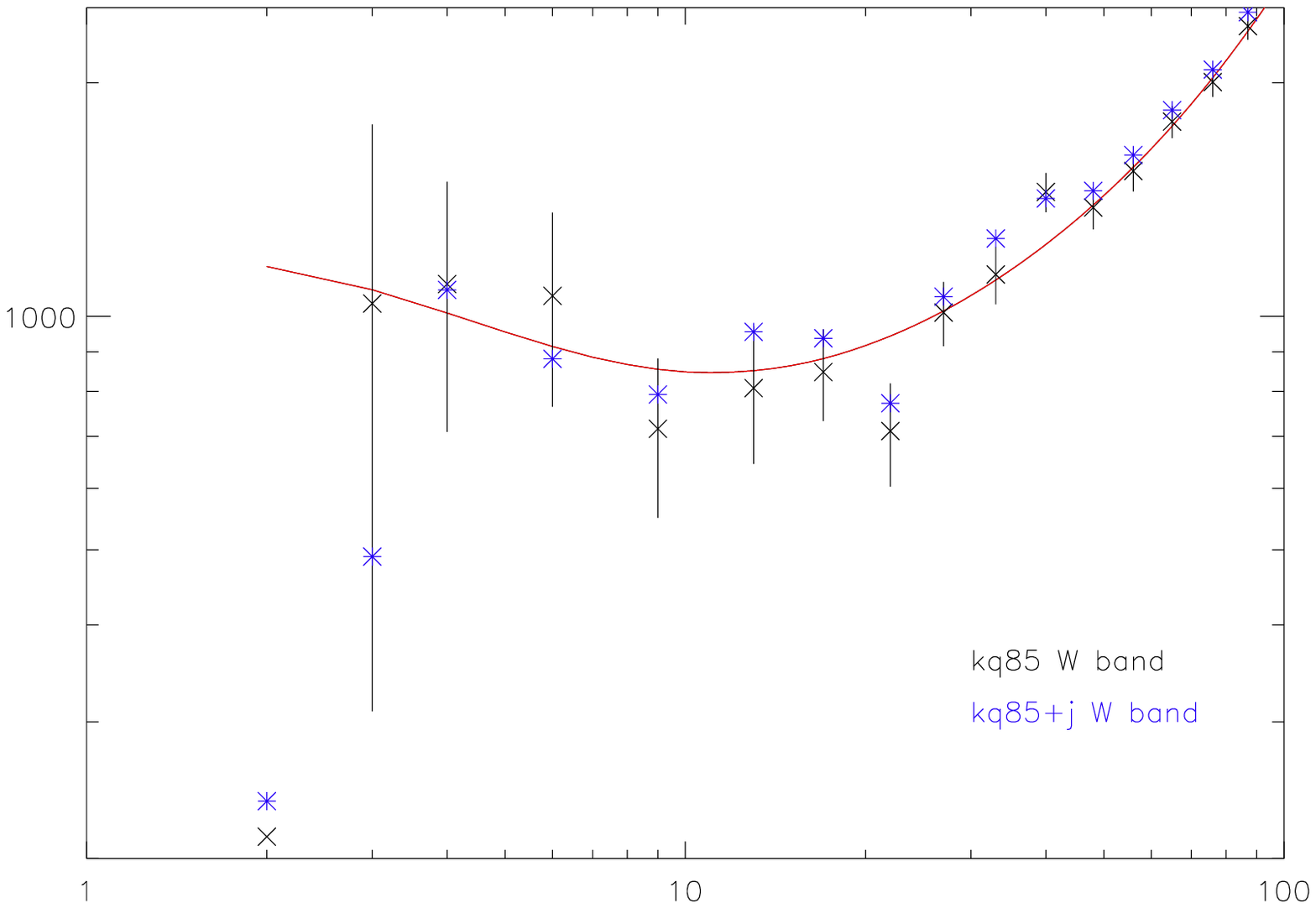}
\caption{Top panel, profile of the function $b(x)$ in $\ell$-space for the choice $B=1.2$. The red dot-dashed line represents $j=17$ and the blue long dashed line $j=20$. Lower panel, power spectrum modification due to the structures measured using the set of needlets shown in Fig.~\ref{fig:j17j20}.}
\label{fig:b1.2profile}
\efg

We find that the effect on the angular power spectrum is actually rather small. However it does follow
the expected sign, decreasing the deficit and the excess of power in the selected ranges.
The effect is shown in the lower panel of Fig.~\ref{fig:b1.2profile}.

%\bfg[htb]
%\incgr[width=\columnwidth]{figures/HLbumps_ps_zoom.eps}
%\caption{Power spectrum modification due to the structures measured using the set of needlets shown in Fig.~\ref{fig:j17j20}.}
%\label{fig:j17j20_spectrum}
%\efg

\section{Conclusions}
\label{sec:conclusions}

We apply spherical needlets to the Wilkinson Microwave Anisotropy Probe 5-year cosmic microwave background
dataset, to  search for imprints of nonisotropic features in the CMB sky.

%HERE I HAVE A CUT A BIT BECAUSE SOME OF THESE THINGS WERE MENTIONED BEFORE,
%AND OTHERS ARE COMMON WITH OTHER KIND OF WAVELETS.

%Spherical needlets have the useful property that features
% can be localized both in harmonic (or multipolar) space as well as in pixel space.
%This allows us both to resolve peculiar features in the CMB sky and to study how these features contribute to the
%anisotropy power spectrum of the CMB.

After calibration by means of a large set of mock simulations, the analysis of needlet coefficients
highlights the presence of the now well-known ``cold spot'' of the CMB map in the southern hemisphere, and in addition
two hot spots at significance greater than  99\% confidence level,
again in the southern hemisphere and closer to the Galactic plane.
While the cold spot primarily contributes to the anisotropy power spectrum  in the
multipoles between $\ell=6$ to $\ell=33$, the hot spots  are found to be dominating the anisotropy power
in the range between $\ell=6$ and $\ell=18$.

We also studied the effect the two spots have on the CMB power spectrum, by building $1000$ mock CMB simulations. We
conclude that, especially at low multipoles, the effect is measurable: masking both the cold and the two hot spots
results in an increase in the quadrupole amplitude of 10\%, while at $\ell=10$ power is reduced by 12\%.
To investigate the effect of this difference on the value of cosmological parameters,
we modified the WMAP 5-year CMB fiducial
power spectrum and the KQ$85$ mask used to perform the analysis by cutting out the contribution of the two spots, and
we repeated the parameter estimation analysis.
The results we obtain are slightly different, but fully consistent within the 1$\sigma$ errors on parameters published
by the WMAP team.

Since all three spots appear in the southern hemisphere, we also studied the power spectrum asymmetry between
the two hemispheres, which has been previously found to be statistically significant.
When the features detected by needlets are masked,
we find that the difference in the power, measured in terms of the anisotropy variance between $\ell=4$ and $\ell=18$, is
reduced by a factor of $2$. This decreases the significance of the previously claimed north-south asymmetry.
We make the mask resulting from needlet features available for future, more detailed studies on
the asymmetries in the CMB anisotropy sky\footnote{http://www.fisica.uniroma2.it/$\sim$cosmo/masks}.

\begin{center}
{\bf Acknowledgments}
\end{center}

%This  work   was  supported  by   NSF  CAREER AST-0645427.
DP thanks Marcella Veneziani for useful discussions and the UCI Physical Sciences Department where this work has been performed. DM is grateful to P.Baldi,
G.Kerkyacharian and D.Picard for many useful discussions.

\newpage

\bibliography{Bcmbanis_need_ext}

\appendix

\end{document}